\documentclass[twocolumn,journal]{IEEEtranTIE}
\usepackage[T1]{fontenc}
\usepackage[latin9]{inputenc}
\usepackage{array}
\usepackage{refstyle}
\usepackage{booktabs}
\usepackage{multirow}
\usepackage{amsmath}
\usepackage{amssymb}
\usepackage{graphicx}
\usepackage[unicode=true,
 bookmarks=true,bookmarksnumbered=true,bookmarksopen=true,bookmarksopenlevel=1,
 breaklinks=false,pdfborder={0 0 0},pdfborderstyle={},backref=false,colorlinks=false]
 {hyperref}
\hypersetup{pdftitle={Your Title},
 pdfauthor={Your Name},
 pdfpagelayout=OneColumn, pdfnewwindow=true, pdfstartview=XYZ, plainpages=false}

\makeatletter

\AtBeginDocument{\providecommand\Secref[1]{\ref{Sec:#1}}}
\AtBeginDocument{\providecommand\Figref[1]{\ref{Fig:#1}}}
\AtBeginDocument{\providecommand\Tabref[1]{\ref{Tab:#1}}}
\AtBeginDocument{\providecommand\Subsecref[1]{\ref{Subsec:#1}}}
\newcommand{\lyxmathsym}[1]{\ifmmode\begingroup\def\b@ld{bold}
  \text{\ifx\math@version\b@ld\bfseries\fi#1}\endgroup\else#1\fi}

\providecommand{\tabularnewline}{\\}
\RS@ifundefined{subsecref}
  {\newref{subsec}{name = \RSsectxt}}
  {}
\RS@ifundefined{thmref}
  {\def\RSthmtxt{theorem~}\newref{thm}{name = \RSthmtxt}}
  {}
\RS@ifundefined{lemref}
  {\def\RSlemtxt{lemma~}\newref{lem}{name = \RSlemtxt}}
  {}

\let\oldforeign@language\foreign@language
\DeclareRobustCommand{\foreign@language}[1]{%
  \lowercase{\oldforeign@language{#1}}}

\usepackage[caption=false,font=footnotesize]{subfig}
\usepackage{graphicx}
\usepackage{cite}
\usepackage{picinpar}
\usepackage{amsmath}
\usepackage{stfloats}
\usepackage{url}
\usepackage{colortbl}
\usepackage{soul}
\usepackage{multirow}
\usepackage{pifont}
\usepackage{color}
\usepackage{alltt}
\usepackage{enumerate}
\usepackage{siunitx}
\usepackage{breakurl}
\usepackage{epstopdf}
\usepackage{pbox}

\makeatother

\begin{document}
\title{A Class D Power Amplifier for Multi-Frequency Eddy Current Testing
Based on Multi-Simultaneous-Frequency Selective Harmonic Elimination
Pulse Width Modulation}
\author{Yang~Tao,~\IEEEmembership{Member,~IEEE,}~Christos Ktistis,~Yifei~Zhao,~Wuliang~Yin,~\IEEEmembership{Senior Member,~IEEE,}~and~Anthony
J.~Peyton\thanks{Manuscript received March 22, 2019; revised July 17, 2019 and September
2, 2019; accepted October 8, 2019.This work was supported by the U.K.
Engineering and Physical Sciences Research Council via the University
of Manchester Impact Accelerator Account under Grant EP/R511626/1.
(Corresponding author: Yang Tao; Wuliang Yin.)}\thanks{The authors are with the Department of Electrical and Electronic Engineering,
School of Engineering, The University of Manchester, Manchester M13
9PL, U.K. (e-mail: yang.tao@manchester.ac.uk; christos. ktistis@manchester.ac.uk;
yifei.zhao@manchester.ac.uk; wuliang.yin@ manchester.ac.uk; a.peyton@manchester.ac.uk).}\thanks{Color versions of one or more of the figures in this article are available
online at http://ieeexplore.ieee.org.}\thanks{Digital Object Identifier 10.1109/TIE.2019.2947842}}
\markboth{IEEE Transactions on Industrial Electronics}{Yang Tao \MakeLowercase{\emph{et al.}}: A Class D Power Amplifier
for Multi-Frequency Eddy Current Testing Based on Multi-Simultaneous-Frequency
Selective Harmonic Elimination Pulse Width Modulation}
\maketitle
\begin{abstract}
Efficiency and multisimultaneous-frequency (MSF) output capability
are two major criteria characterizing the performance of a power amplifier
in the application of multifrequency eddy current testing (MECT).
Switch-mode power amplifiers are known to have a very high efficiency,
yet they have rarely been adopted in the instrumental development
of MECT. In addition, switch-mode power amplifiers themselves are
lacking in the research literature for MSF capability. In this article,
a Class D power amplifier is designed so as to address the two issues.
An MSF selective harmonic elimination pulsewidth modulation method
is proposed to generate alternating magnetic fields, which are rich
in selected harmonics. A field-programmable-gate-array-based experimental
system has been developed to verify the design. Results show that
the proposed methodology is capable of generating high MSF currents
in the transmitting coil with a low distortion of signal.
\end{abstract}

\begin{IEEEkeywords}
Class D power amplifier, eddy current, multifrequency, pulsewidth
modulation (PWM), selective harmonic elimination.
\end{IEEEkeywords}

\section{Introduction}

\IEEEPARstart{M}{ulti-frequency} eddy current testing (MECT) has
attracted increasing attention for non-destructive testing (NDT) in
recent years. In comparison with the single-frequency eddy current
method, MECT takes advantage of wide-band spectral signals, which
is advantageous for applications such as detecting cracks in conductive
materials \cite{bernieri2013crackdepth}, measuring thickness of metal
film \cite{yin2007thickness}, monitoring microstructure in steel
production \cite{johnstone2001usingelectromagnetic} and imaging conductivity
profile of metal structures \cite{wuliangyin2005imaging}, etc. In
these applications of MECT, a high-current power amplifier with multi-simultaneous-frequency
(MSF) output capability is indispensable in the signal chain so as
to achieve high overall sensitivity \cite{garcia-martin2011nondestructive}.

Class D power amplifier has the virtue of high current handling capacity
and power efficiency \cite{rodriguez2002multilevel,malinowski2010asurvey}.
However, transmitting signals may suffer severe distortion after amplification
if they are not modulated properly. A notable category of modulation
methods is known as the programmed pulse width modulation (PPWM) or
selective harmonic elimination pulse width modulation (SHEPWM) \cite{dahidah2015areview}
which has drawn tremendous interests and been studied primarily for
high-power high-voltage converters \cite{napoles2010selective}. Owing
to the lack of motivation of generating multi-simultaneous frequencies
using SHEPWM, efforts have been mainly devoted in a manner to achieve
varying outputs of the fundamental component which is characterised
by the modulation index and to eliminate the remaining harmonics.
Yet in MECT applications, the varying modulation index is not of interest
while the MSF output capability is of importance. Therefore, some
harmonics may be well harnessed rather than eliminated.

This paper attempts, for the first time, to examine the MSF capability
of the SHEPWM technique and in addition, to introduce switch-mode
power amplifier into the instrumental development of MECT. A Class
D power amplifier based upon multi-simultaneous-frequency SHEPWM (MSF-SHEPWM)
is designed. The proposed design is capable of driving an excitation
coil with a simultaneous multi-frequency signal. Moreover, it is convenient
to change the spectrum of the transmitting signal. Experimental results
have proved that the system could transmit MSF alternating magnetic
fields with low distortion. The remainder of the paper is organised
as follows. \Secref{MSFPWM-Principle} describes the principle of
MSF-SHEPWM. The design of hardware and the results of experiments
are presented in \Secref{Hardware-Design}. Finally, conclusions are
drawn in the last section.

\section{MSF-SHEPWM Principle\label{sec:MSFPWM-Principle}}

\begin{figure}[tbh]
\begin{centering}
\includegraphics[width=1\columnwidth]{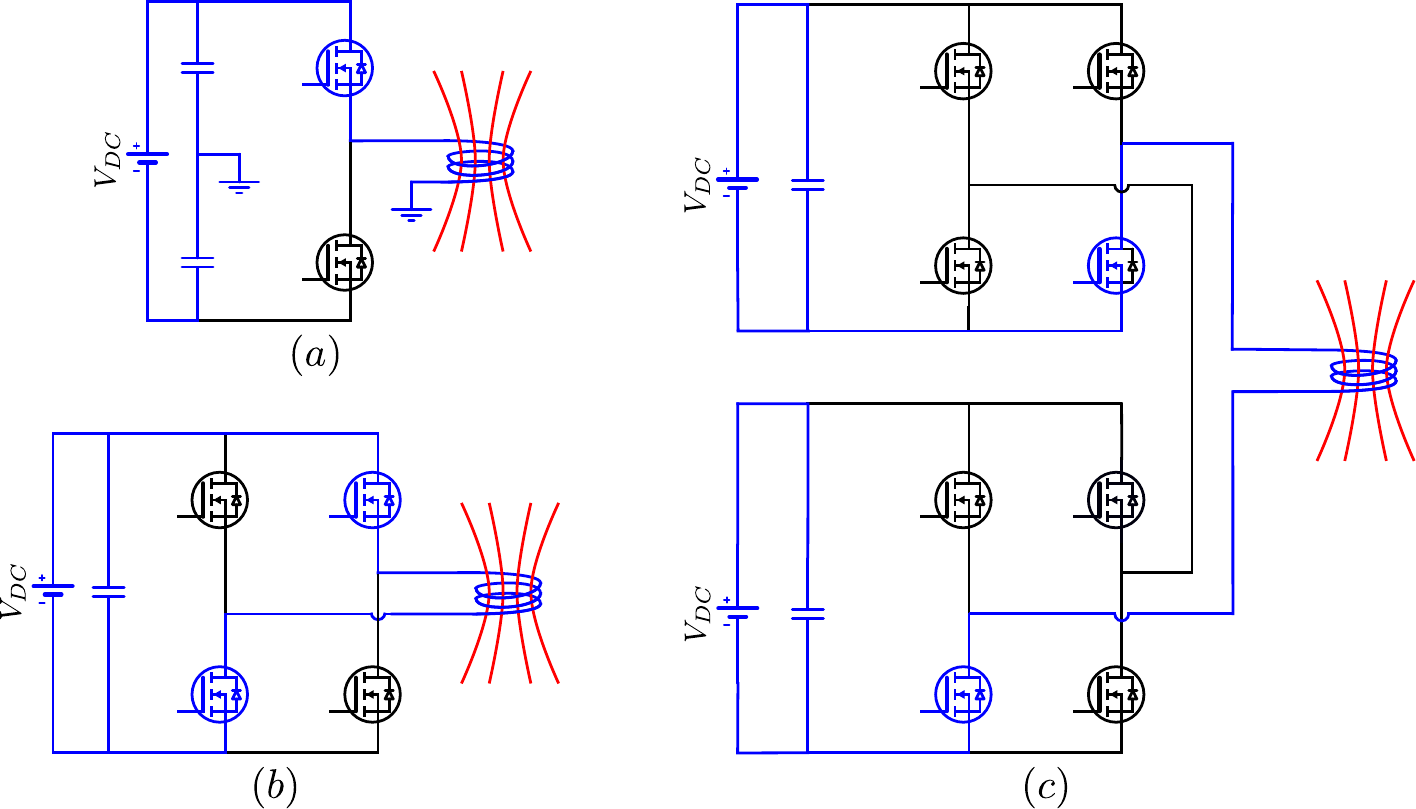}
\par\end{centering}
\caption{Topology of Class D power amplifier using enhancement-mode n-type
MOSFETs. (a) push-pull half bridge. (b) full H-bridge. (c) two full
H-bridges.\label{fig:Topology-of-Class}}
\end{figure}

In eddy current testing, the load of the power amplifier is a coil
that generates alternating magnetic fields. In order to generate the
magnetic fields that are required for MECT, according to the $\mathrm{Amp\grave{e}re's}$
law, we can directly design the current through the coil. The waveform
of coil current is a concatenation of a few straight line segments
of fixed gradients when the coil is connected to a Class D power amplifier.
This is due to the fact that in theory the voltage across the coil
can only have a fixed number of values or levels and, for a perfect
inductor, the coil current is an integral of the voltage. \Figref{Topology-of-Class}
shows three example topologies for Class D power amplifier. Enhancement-mode
n-type MOSFETs are used for this illustration. The components in blue
colour constitute a conducting path. \Figref{Topology-of-Class}(a)
shows a push-pull configuration where the voltage across the coil
have two levels, i.e. $+V_{DC}/2$ and $-V_{DC}/2$; \Figref{Topology-of-Class}(b)
depicts a full H-bridge topology that is capable of driving the coil
with three-level voltage, i.e. $+V_{DC}$, $0$ and $-V_{DC}$. Multi-level
voltage can be achieved by utilising multiple full H-bridges. For
example, a five-level inverter is shown in \Figref{Topology-of-Class}(c)
that contains two H-bridge cells. The five values of voltage are $+2V_{DC}$,
$+V_{DC}$, $0$, $-V_{DC}$ and $-2V_{DC}$.

The objective of MSF-SHEPWM is to generate an output waveform that
contains the desired MSF spectral components, but eliminates the unwanted
harmonics. The principle of MSF-SHEPWM can be summarised in three
stages. Firstly, the objective signal is synthesised based on the
fundamental sinusoidal signal and a few harmonics in the time domain,
and the waveform of the synthesised signal is resembled by straight
line segments in a morphological sense. Secondly, the resultant resemblance
signal is decomposed as Fourier series and analysed in the frequency
domain. Finally, the Fourier coefficients are adjusted by solving
an optimisation problem. These three stages will now be described
in detail in the following sub-sections. For clarity we use $f_{k}(t)$
to denote the objective signal, $g_{k}(t)$ as the resemblance function
of $f_{k}(t)$, and $G_{k}(t)$ is the Fourier series of $g_{k}(t)$.
The subscript $k$ refers to a specific signal.

\subsection{Synthesis of the objective function $f_{k}(t)$ in the time domain\label{subsec:Synthesis-of-harmonics}}

A periodic signal $f(t)$ that contains $n$ sinusoidal basic components
is expressed as (\ref{eq:f(t)}) where $\omega_{p}$, $\theta_{p}$
and $w_{p}$ are the angular frequency, phase angle and coefficient
of the $p^{th}$ component respectively. The function $f(t)$ serves
as the objective coil current that is to be generated and the coefficient
$w_{p}$ forms the amplitude spectrum.

Additionally, three conditions are exerted on $f(t)$. Firstly, the
lowest frequency is set as the fundamental frequency $\omega$ and
higher frequencies are harmonics of $\omega$, i.e. $\omega_{p}=p\omega$.
This is a natural requirement since the signal generated by Class
D power amplifier is rich in harmonics. Secondly, the phase angle
$\theta_{p}$ is zero implying all the basic components are in-phase.
In MECT applications, the relative phase values are not critical factors.
Thirdly, the coefficient $w_{p}$ is chosen as $\frac{1}{p}$ for
the selected frequencies and zero for absent harmonics. In general,
the magnitude of response signal tends to increase with frequency.
In order to achieve similar signal-to-noise ratio (SNR) in the whole
spectrum of interest, the coefficient $w_{p}$ is chosen to be inversely
proportional to the frequency $\omega_{p}$ or the order $p$. In
fact, (\ref{eq:f(t)}) is a special case of the trigonometric form
of Fourier series but only a few rather than an infinite number of
components are involved.

\begin{equation}
f(t)=\sum_{p=1}^{n}w_{p}\sin\left(\omega_{p}t+\theta_{p}\right)\label{eq:f(t)}
\end{equation}
\begin{multline}
f_{1}(t)=\sin\left(\omega t\right)+\frac{1}{3}\sin\left(3\omega t\right)+\frac{1}{9}\sin\left(9\omega t\right)\\
+\frac{1}{27}\sin\left(27\omega t\right)+\frac{1}{81}\sin\left(81\omega t\right)\label{eq:f1}
\end{multline}

\begin{multline}
f_{2}(t)=\sin\left(\omega t\right)+\frac{1}{3}\sin\left(3\omega t\right)+\frac{1}{7}\sin\left(7\omega t\right)\\
+\frac{1}{17}\sin\left(17\omega t\right)\label{eq:f2}
\end{multline}

\begin{multline}
f_{3}(t)=\sin\left(\omega t\right)+\frac{1}{2}\sin\left(2\omega t\right)+\frac{1}{4}\sin\left(4\omega t\right)\\
+\frac{1}{6}\sin\left(6\omega t\right)+\frac{1}{8}\sin\left(8\omega t\right)+\frac{1}{10}\sin\left(10\omega t\right)\\
+\frac{1}{12}\sin\left(12\omega t\right)+\frac{1}{14}\sin\left(14\omega t\right)\label{eq:f3}
\end{multline}

\begin{multline}
f_{4}(t)=\sin\left(\omega t\right)+\frac{1}{2}\sin\left(2\omega t\right)+\frac{1}{3}\sin\left(3\omega t\right)\\
+\frac{1}{4}\sin\left(4\omega t\right)+\frac{1}{5}\sin\left(5\omega t\right)+\frac{1}{6}\sin\left(6\omega t\right)\\
+\frac{1}{7}\sin\left(7\omega t\right)+\frac{1}{8}\sin\left(8\omega t\right)\label{eq:f4}
\end{multline}

\begin{multline}
f_{5}(t)=\sin\left(\omega t\right)+\frac{1}{3}\sin\left(3\omega t+\frac{\pi}{4}\right)+\frac{1}{9}\sin\left(9\omega t+\frac{\pi}{7}\right)\\
+\frac{1}{27}\sin\left(27\omega t+\frac{\pi}{5}\right)+\frac{1}{81}\sin\left(81\omega t+\frac{\pi}{6}\right)\label{eq:f5}
\end{multline}

\begin{figure}[tbh]
\begin{centering}
\includegraphics[width=0.9\columnwidth]{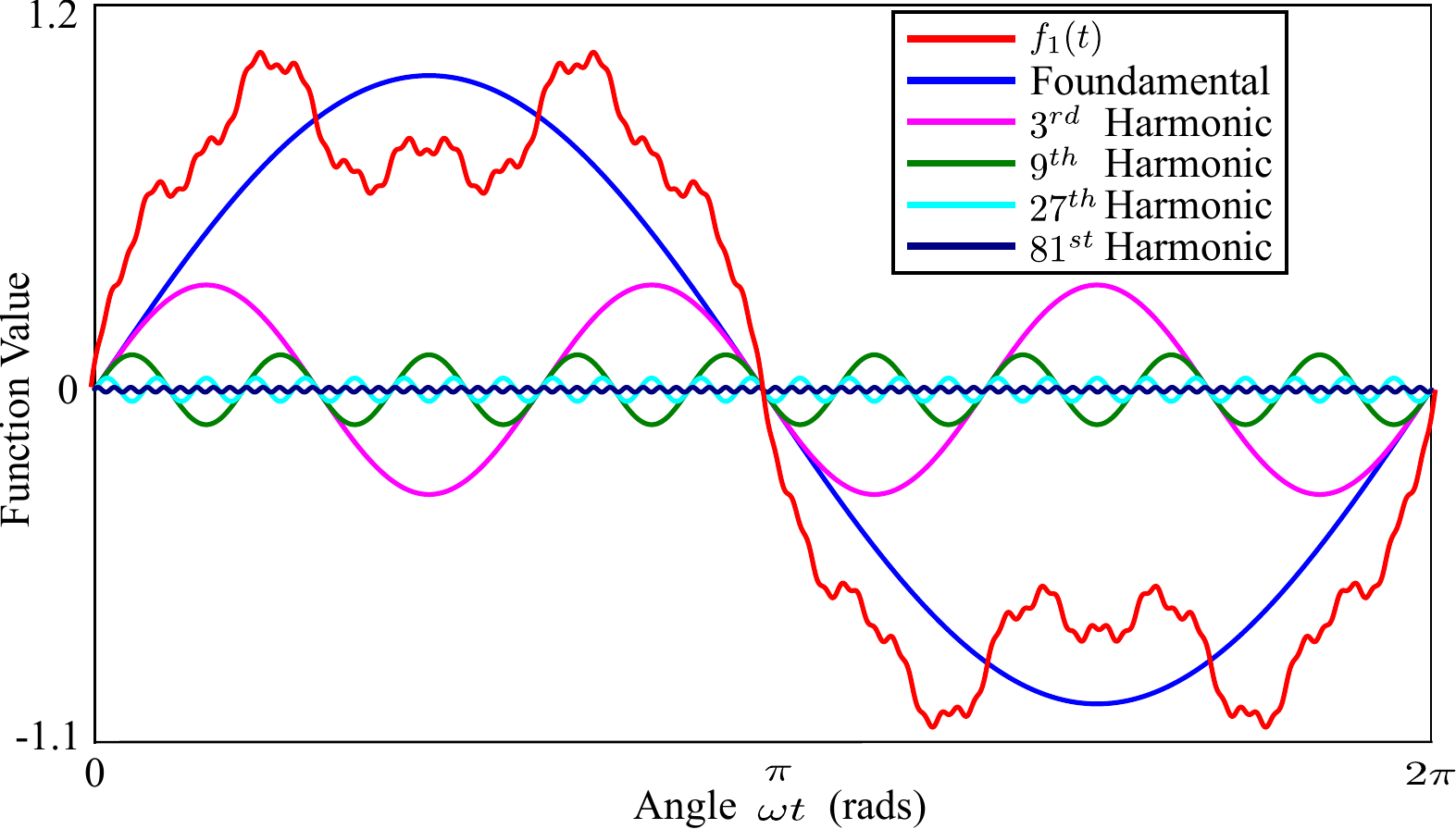}
\par\end{centering}
\caption{The objective function $f_{1}(t)$, the corresponding fundamental
and harmonic functions\label{fig:Plots-of-signal}}
\end{figure}

\begin{figure}[tbh]
\begin{centering}
\includegraphics[width=0.9\columnwidth]{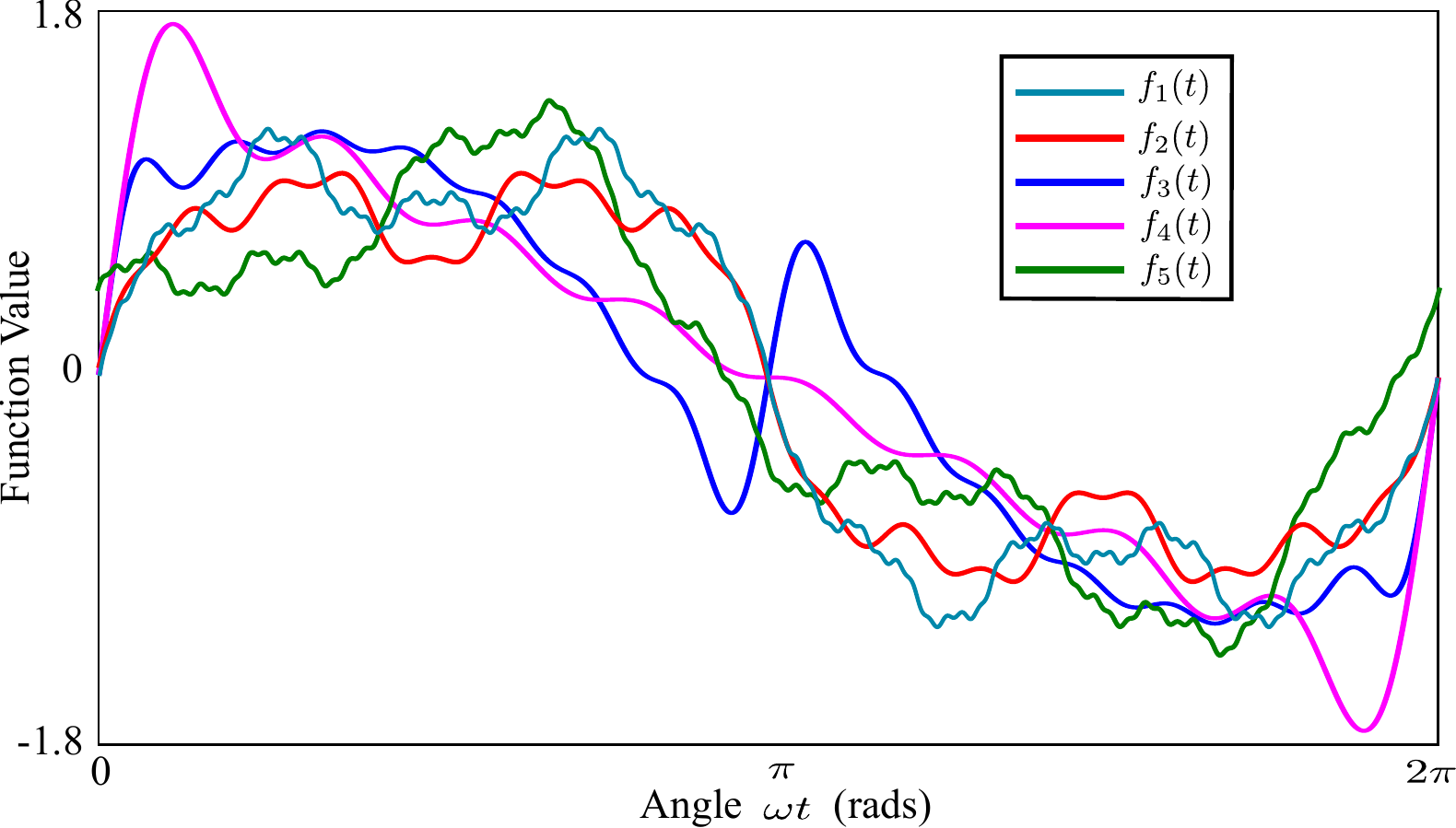}
\par\end{centering}
\caption{Objective function: $f_{1}(t)$, $f_{2}(t)$, $f_{3}(t)$, $f_{4}(t)$
and $f_{5}(t)$ \label{fig:Examples}}
\end{figure}

As an example, a signal $f_{1}(t)$ that follows these principles
is shown in \Figref{Plots-of-signal}. Here, $f_{1}(t)$ is a superposition
of five basic sinusoidal signals. The frequencies are chosen in a
geometric scale as $\omega$, $3\omega$, $9\omega$, $27\omega$
and $81\omega$. The weights are set as $1$, $\frac{1}{3}$, $\frac{1}{9}$,
$\frac{1}{27}$, and $\frac{1}{81}$. Another three examples $f_{2}(t)$,
$f_{3}(t)$ and $f_{4}(t)$ that have different possible combinations
of basic components are shown in \Figref{Examples}. The expressions
of $f_{1}(t)$, $f_{2}(t)$, $f_{3}(t)$ and $f_{4}(t)$ are shown
in (\ref{eq:f1}), (\ref{eq:f2}), (\ref{eq:f3}) and (\ref{eq:f4}),
respectively.

The selection of involving harmonics has flexibility and is determined
by the fundamental frequency and the spectrum of interest. The associated
coefficients are determined by the requirement of power distribution
for each harmonics. It should be noted that the basic sinusoidal components
do not necessarily have to be in-phase relative to each other. For
example, $f_{5}(t)$ as expressed in (\ref{eq:f5}), has the same
weights for each basic components as $f_{1}(t)$, but the phases are
different which is shown in \Figref{Examples}. Although $f_{5}(t)$
has the same amplitude spectrum as $f_{1}(t)$, some symmetric features
are broken which makes the analysis in the frequency domain more complicated.

\begin{figure}[tbh]
\begin{centering}
\includegraphics[width=0.9\columnwidth]{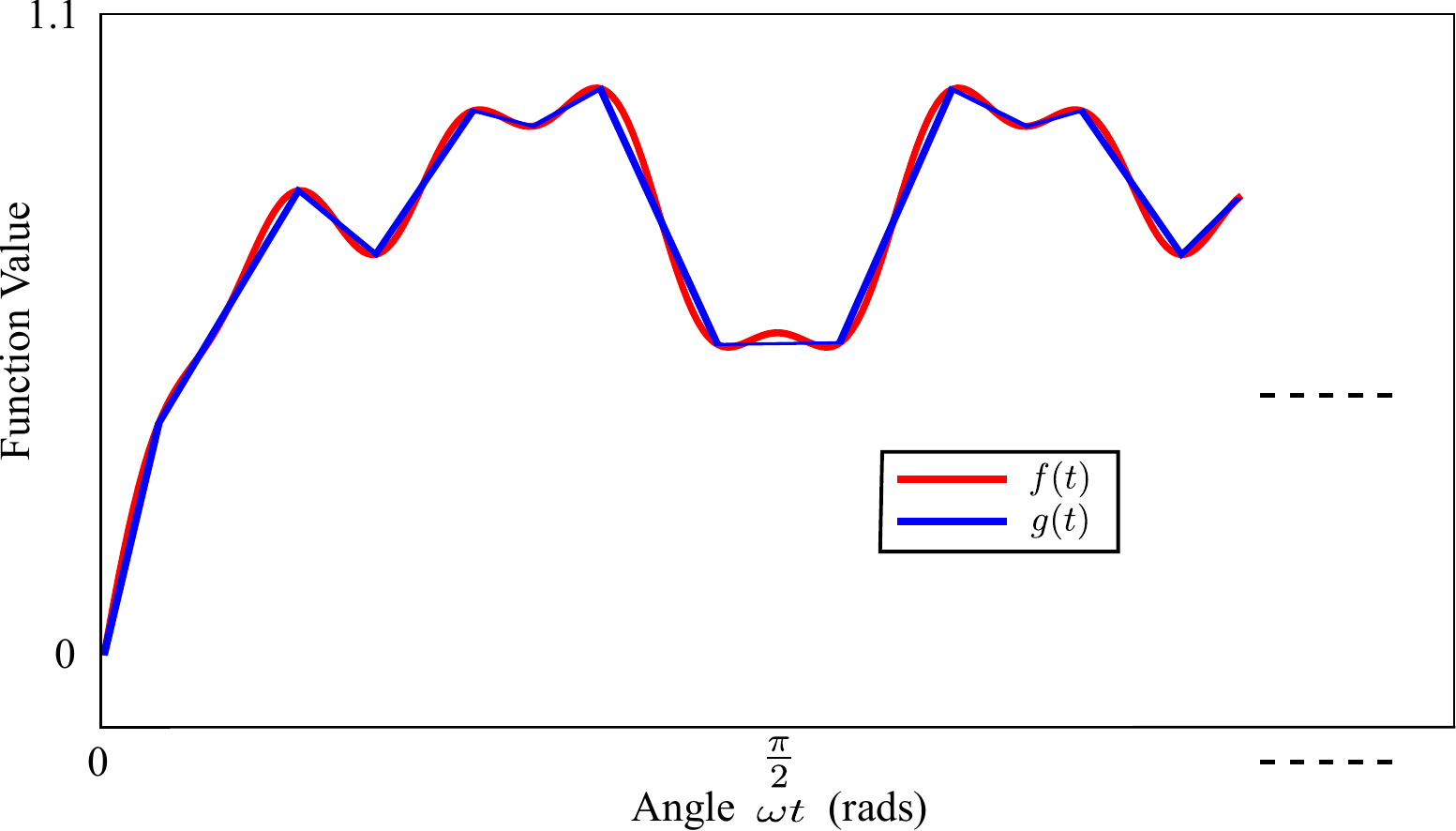}
\par\end{centering}
\caption{An objective function $f(t)$ and its resemblance function $g(t)$\label{fig:The-objective-function}}
\end{figure}

Once the waveform of the objective function is determined, such waveform
can be resembled using straight line segments as shown in \Figref{The-objective-function}.
The resemblance function $g(t)$ is piece-wise linear with different
gradients and approximates the objective function $f(t)$. Accurate
approximation requires multiple levels of output voltage which, however,
costs multiple H-bridge cells. Therefore, the selection of voltage
levels is a trade-off between the fidelity of signal and feasibility
in practice. The number of time steps used for $g(t)$ is often determined
by the clock frequency used to drive the bridge circuit and limited
by the dynamic characteristics of the transistor.

\subsection{Analysis of the resemblance function $g_{k}(t)$ in the frequency
domain}

\subsubsection{Calculation of the Fourier coefficients for the resemblance function
$g_{k}(t)$}

\begin{figure}[tbh]
\begin{centering}
\includegraphics[width=1\columnwidth]{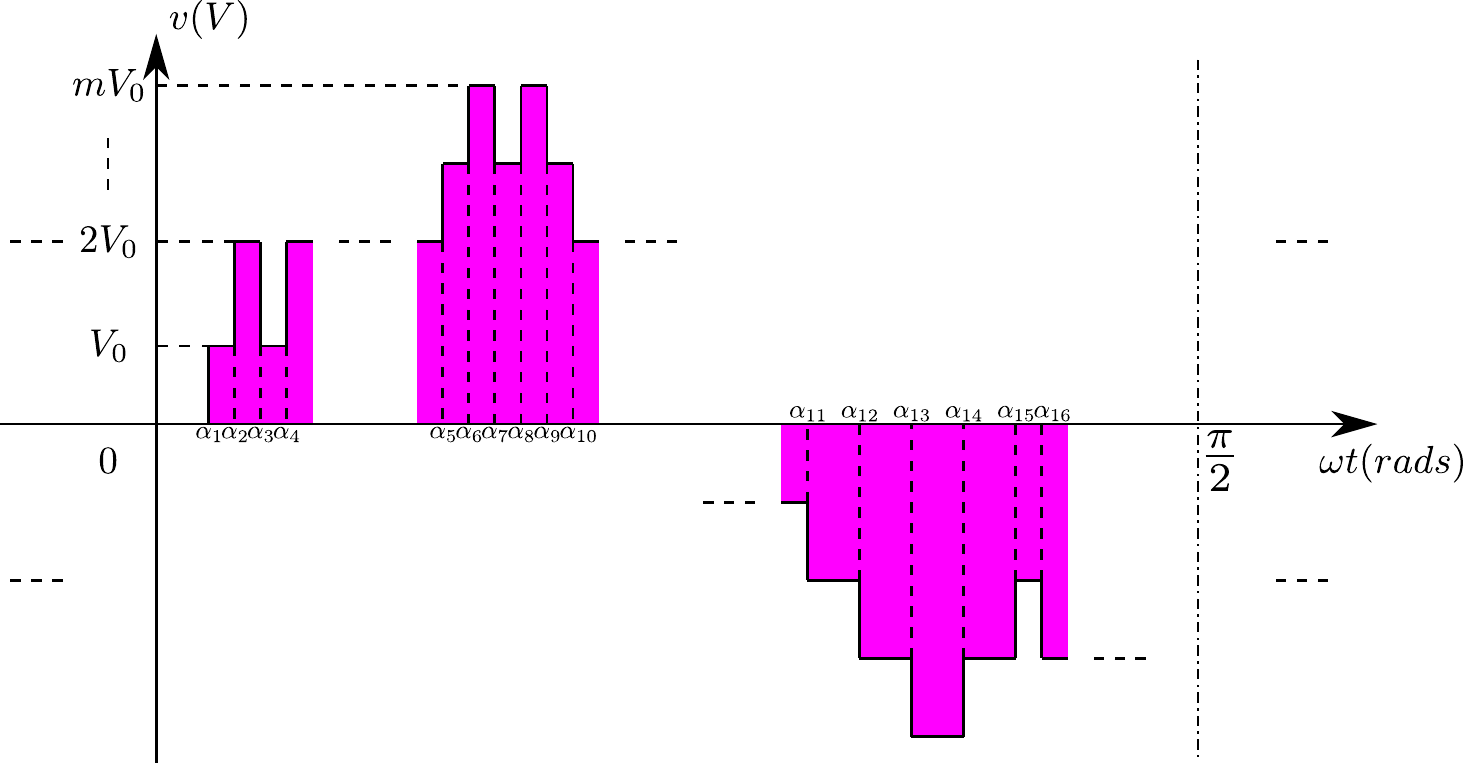}
\par\end{centering}
\caption{Switching scheme of multi-level Class D amplifier\label{fig:Switching-scheme-of}}
\end{figure}

A resemblance function $g_{k}(t)$ can be generated by a multi-level
Class D power amplifier. The switching scheme in a quarter period
is depicted in \Figref{Switching-scheme-of}. Switching angles satisfy
$0\leq\alpha_{1}\leq\alpha_{2}\leq\cdots\leq\alpha_{N}\leq\frac{\pi}{2}$
and at each angle the voltage can raise or fall by $V_{0}$ until
reaching the limits. The coil current is essentially an integral of
the pulse voltage which is depicted as the area in the magenta colour
in \Figref{Switching-scheme-of}. Therefore, the Fourier series of
the coil current is calculated as (\ref{eq:fft-multi}). The Fourier
coefficient is shown in (\ref{eq:fft-multi-coefficient}) which is
a function of $N$ switching angles $\alpha_{q}$, the coil inductance
$L$, the order of harmonics $p$ and the step of voltage $V_{0}$.

\begin{equation}
G(t)=\sum_{p=1,3,\cdots}^{\infty}b_{p}\sin\left(p\omega t\right)\label{eq:fft-multi}
\end{equation}

\begin{align}
b_{p} & =-\frac{4V_{{\it 0}}}{p^{2}\pi\omega L}\sum_{q=1}^{N}(-1)^{s}\sin\left(p\alpha_{q}\right),\nonumber \\
s & =\begin{cases}
1 & ,\forall\,\alpha_{q}\,at\,which\,there\,is\,a\,falling\,edge\\
0 & ,\forall\,\alpha_{q}\,at\,which\,there\,is\,a\,raising\,edge
\end{cases}\label{eq:fft-multi-coefficient}
\end{align}

\subsection{Optimisation of the Fourier coefficients $b_{p}$}

Once the waveform of the resemblance function $g_{k}(t)$ is obtained
based on the gradient of $f_{k}(t)$, the spectrum of $g_{k}(t)$
can be controlled by adjusting the switching angles $\alpha_{q}$
which is essentially a problem of solving simultaneous transcendental
equations. Each equation corresponds to the magnitude of a harmonic
to be controlled.

Various methods have been developed in order to obtain one or multiple
solutions to the harmonic magnitude equations. One direction for developing
a solver is based on the elimination theory of polynomial and using
the $resultant$ \cite{chiasson2003control}. Unfortunately, the order
of polynomials increases with the number of harmonics to be controlled
which results in a significant computational burden if the application
entails a wide spectrum. Moreover, completely solving the equations
might be a stringent requirement and in fact discrepancies could be
tolerated within a threshold in a real application. Another direction
tends to recast the harmonic magnitude equations into an optimisation
problem \cite{dahidah2015areview}. Because of the trigonometric and
non-convex nature of the harmonic magnitude equations, it is difficult
to guarantee that a globally optimal solution can be achieved. Meta-heuristic
algorithms such as differential evolution \cite{amjad2015application},
genetic algorithm \cite{lee2016switchedbattery}, and particle swarm
optimisation \cite{taghizadeh2010harmonic}, etc. have been applied
to obtain a global solution through modern stochastic search techniques.
However, different results may be returned for the same problem since
these algorithms rely on random searches. Many gradient based algorithms,
on the other hand, guarantee that a local optimal can be obtained
in theory. If a good estimation of the solution exists, the gradient
based algorithms could converge to the optimal rapidly \cite{boyd2004convexoptimization}.

\begin{multline}
\min_{0\leq\alpha_{1}\leq\alpha_{2}\leq\cdots\leq\alpha_{N}\leq\frac{\pi}{2},m_{p}}\sum_{p\in\Phi}\lambda_{p}\left(\frac{p\omega L}{V_{0}}b_{p}-m_{p}\right)^{2},\\
subject\,to\,\left|\frac{p\omega L}{V_{0}}b_{p}\right|\leq\epsilon_{p},\,\forall p\in\Psi\label{eq:opt}
\end{multline}

In this paper, the harmonic magnitude equations are calculated by
solving a constrained nonlinear optimisation problem as shown in (\ref{eq:opt}).
Following the convention of multi-level inverter, we denote the modulation
index for the $p^{th}$ harmonic frequency as $m_{p}=\frac{V_{p}}{V_{0}}$,
where $V_{p}$ is the amplitude of the $p^{th}$ harmonic in the output
voltage. The modulation index characterises the relative amplitude
of output voltage in the context of a multi-level inverter, but there
are two different considerations for MECT. Firstly, $m_{p}$ is of
interest for multiple harmonics, yet only the fundamental frequency
component is relevant for a multi-level inverter. Secondly, $m_{p}$
is treated as an optimising parameter, while a multi-level inverter
tends to be able to output a range of values of the modulation index.
In (\ref{eq:opt}), $\Phi$ is a set whose elements are the orders
of selected harmonics while the elements of $\Psi$ are the orders
of harmonics to be eliminated. The objective function in (\ref{eq:opt})
represents the sum of weighted ($\lambda_{p}$) squared residuals
of the modulation index for selected harmonics subject to such constrains
that the modulation index of undesirable harmonics is bounded with
threshold $\epsilon_{p}$. The optimising parameters are $N$ switching
angles $\alpha_{1},\alpha_{2},\cdots,\alpha_{N}$ and the modulation
index $m_{p}$.

The problem (\ref{eq:opt}) could provide many flexibilities in practice.
For instance, the adjacent harmonics of an objective frequency will
cause interference at that frequency and hence the corresponding threshold
$\epsilon_{p}$ for the adjacent harmonics should be smaller than
those of other less critical harmonics. Additionally, if annulation
of the objective function is not possible, the objective function
could be relaxed by putting higher weights $\lambda_{p}$ for more
important harmonics and lower weights for other harmonics.

The interior-point algorithm is adopted to solve the problem (\ref{eq:opt}).
This choice of algorithm is due to the fact that the interior-point
algorithm has been proved to be an effective solver for constrained
nonlinear optimisation problems \cite{nesterov1994interiorpoint}.
Moreover, both the objective function and constrain function have
well-defined Gradient and Hessian. Besides, the initial value of switching
angles can be obtained based on the gradient of objective function
$f(t)$ which turns out to be a good estimation. The $MATLAB^{\circledR}$
$Optimization\,Toolbox^{\circledR}$ provides a solver $\mathrm{fmincon}$
that implements the interior-point algorithm.

\section{Hardware and Experimental Results\label{sec:Hardware-Design}}

\subsection{Hardware Implementation}

\begin{figure}[tbh]
\begin{centering}
\includegraphics[width=0.9\columnwidth]{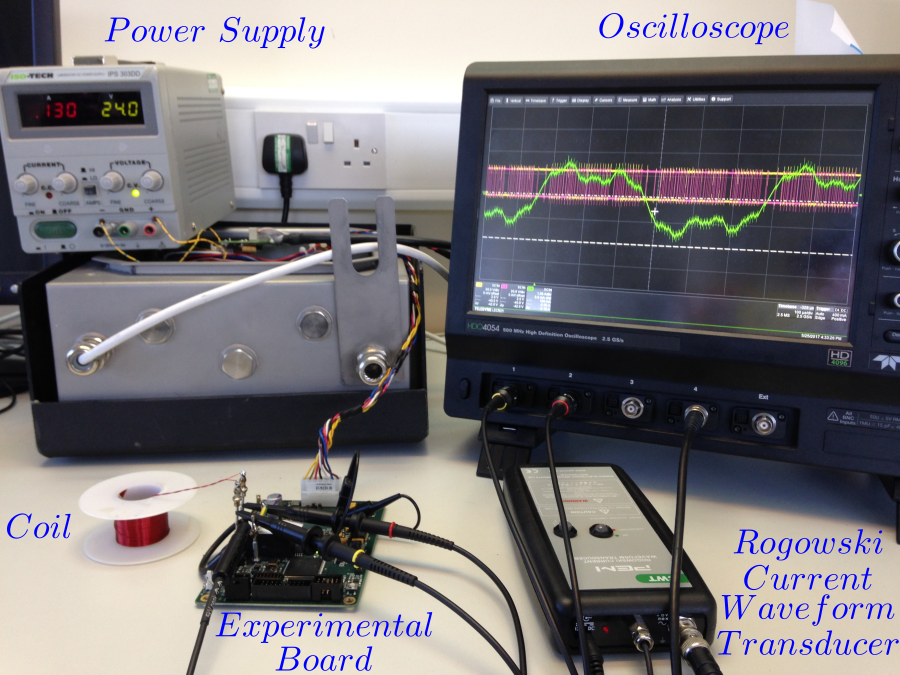}
\par\end{centering}
\caption{Experimental system\label{fig:Experimental-system}}
\end{figure}

\begin{figure}[tbh]
\begin{centering}
\includegraphics[width=0.9\columnwidth]{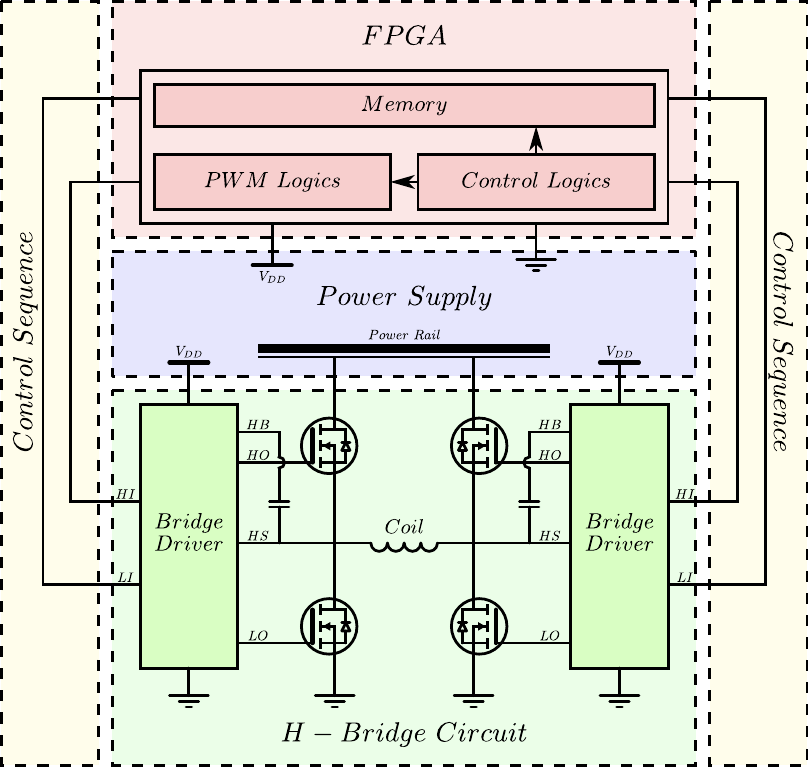}
\par\end{centering}
\caption{Schematic diagram of the experimental system\label{fig:Schematic-diagram-of}}
\end{figure}

In order to test the proposed MSF-SHEPWM method, an FPGA based experimental
system was designed which is shown in \Figref{Experimental-system}.
The system schematic diagram is shown in \Figref{Schematic-diagram-of}.
The system mainly comprises three parts, namely a $Spartan^{\circledR}-6$
FPGA, power supply and H-bridge circuit. An oscillator on the experiment
board generates a clock signal of a fixed frequency, and a digital
clock manager (DCM) inside the FPGA is used to generate a specified
baseline clock signal for control sequence. The FPGA periodically
reads the switching sequence from in-chip memory and outputs pulse
signals through a PWM logic block. The pulse signals are then fed
into the bridge driver. Four enhancement-mode n-type MOSFETs are utilised
in the system configured in a full H-bridge. Therefore, three levels
of voltage can be generated by the H-bridge cell. According to the
analysis of two cases of application that will be shown in the following
paragraphs, even a bipolar pulse voltage is sufficient in generating
the desired magnetic fields. Although the proposed MSF-SHEPWM method
applies to the design of multiple H-bridge cells, only a single full
H-bridge is adopted in the system. In order to make the enhancement-mode
n-type MOSFETs working in the ohmic region, the gate-to-source voltage
must be higher than not only the threshold voltage, but the addition
of the drain-to-source voltage and the threshold voltage. Therefore,
a bootstrap circuit is required to drive the high side MOSFETs. The
dc voltage is selected as $24\,V$ supplied externally.

\subsection{Power and Efficiency Metrics}

The drain efficiency $\eta$ \cite{eroglu2015introduction} as expressed
in (\ref{eq:drainefficiency}) is adopted as an efficiency metric,
in which the input dc power $P_{DC}$ is the product of the dc power
supply voltage $V_{DC}$ and current $I_{DC}$. The real output power
$P_{out}$ is defined in (\ref{eq:pout}), in which $u_{out}$, $i_{out}$
and $T$ are the load voltage, current and integration interval, respectively.
As the objective is generating ac magnetic field rather than dissipating
dc power in the load, we define an energy conversion factor $\zeta$
as (\ref{eq:engcvfctr}), in which the output reactive power $Q_{out}$
\cite{svensson1999powermeasurement} is defined as (\ref{eq:qout})
where $U_{i}$ and $I_{i}$ are the root-mean-squared values of the
output voltage and current harmonics of order $i$. $\theta_{i}$
is the phase angle difference between them, and $\Psi$ is the set
containing the orders of transmitting harmonics.

\begin{equation}
\eta=\frac{P_{out}}{P_{DC}}\label{eq:drainefficiency}
\end{equation}

\begin{equation}
P_{out}=\frac{1}{T}\int_{T}u_{out}i_{out}\;\mathrm{d}t\label{eq:pout}
\end{equation}

\begin{equation}
\zeta=\frac{Q_{out}}{P_{DC}}\label{eq:engcvfctr}
\end{equation}

\begin{equation}
Q_{out}=\sum_{i\in\Psi}U_{i}I_{i}\sin\theta_{i}\label{eq:qout}
\end{equation}

\subsection{Two Experiments}

The selection of excitation frequencies is of high importance and
it can be expected that different spectrum is required for different
applications. For example, the $\beta-dispersion$ frequency range
from hundreds of kilohertz to megahertz is often harnessed to detect
the property which relates to the cellular structure of biological
tissues \cite{otoole2015noncontact}, while $20\,kHz$ to $100\,kHz$
is a suitable range for measuring copper film with thickness of $10\,\lyxmathsym{\textendash}\,400\,\mu m$
\cite{li2017athickness}.

Two experiments are carried out based on the experimental system.
In the first experiment, the full design procedures are illustrated
and the questions needing attentions are discussed. In the second
experiment, we show a fast design without the procedure of analysing
in the frequency domain and the optimisation process. The setups of
two experiments can be found in \Tabref{Details-of-setup}.

\begin{table}[tbh]
\caption{Detailed setups of two experiments\label{tab:Details-of-setup}}

\resizebox{\columnwidth}{!}{
\begin{centering}
\begin{tabular}{c>{\centering}p{0.4\columnwidth}>{\centering}p{0.5\columnwidth}}
\toprule 
 & Experiment No. 1 & Experiment No. 2\tabularnewline
\midrule 
Selected Harmonics & $1^{st}$, $3^{rd}$, $7^{th}$, $17^{th}$ & $1^{st}$, $3^{rd}$, $9^{th}$, $27^{th}$, $81^{st}$\tabularnewline
Clock Frequency & $24\,MHz$ & $6.25\,MHz$\tabularnewline
Lookup Table Length & $476\,bits$ & $3888\,bits$\tabularnewline
Transmitting Frequency & $50.42\,kHz$, $151.26\,kHz$, $352.94\,kHz$, $857.14\,kHz$ & $1.61\,kHz$, $4.82\,kHz$, $14.47\,kHz$, $43.40\,kHz$, $130.21\,kHz$\tabularnewline
Coil Inductance & $1.4\,\mu H$ & $662\,\mu H$\tabularnewline
Wire Diameter (AWG) & $13\,AWG$ & $26\,AWG$\tabularnewline
\bottomrule
\end{tabular}
\par\end{centering}
}
\end{table}

\subsubsection{Experiment No. 1}

The objective of this experiment is to generate high currents for
a low-impedance coil in the frequency range between $10\,kHz$ and
$1\,MHz$. Test samples can be transferred through the aperture of
the coil. This configuration could be applied to inline discrimination
of metal contaminants, grading of food product and monitoring of meat
ageing process, etc.

\paragraph{Excitation frequencies}

Deciding the number of transmitting frequencies is a trade-off between
power distribution and spectral resolution. If too many frequencies
are selected, it's inevitable that the power will be relatively low
for an individual frequency. Moreover, if two selected frequencies
are too close, not enough distinction of the response signal could
be drawn from them. As a result, four frequencies are chosen for this
experiment as the fundamental, $3^{rd}$, $7^{th}$ and $17^{th}$
harmonics which results in the earlier mentioned objective function
$f_{2}(t)$ in \Subsecref{Synthesis-of-harmonics}. The clock frequency
is configured as $24\,MHz$ and the highest frequency is $\frac{1}{28}$
of the clock frequency. Therefore, the four transmitting frequencies
are about $50.42\,kHz$, $151.26\,kHz$, $352.94\,kHz$ and $857.14\,kHz$.
It is noted that these frequencies well span the objective spectrum
but are relatively arbitrary and the transmitting frequencies could
be altered by either changing the clock frequency or the ratio of
the highest frequency over the clock frequency.

\paragraph{Levels of output voltage}

\begin{figure}[tbh]
\begin{centering}
\includegraphics[width=1\columnwidth]{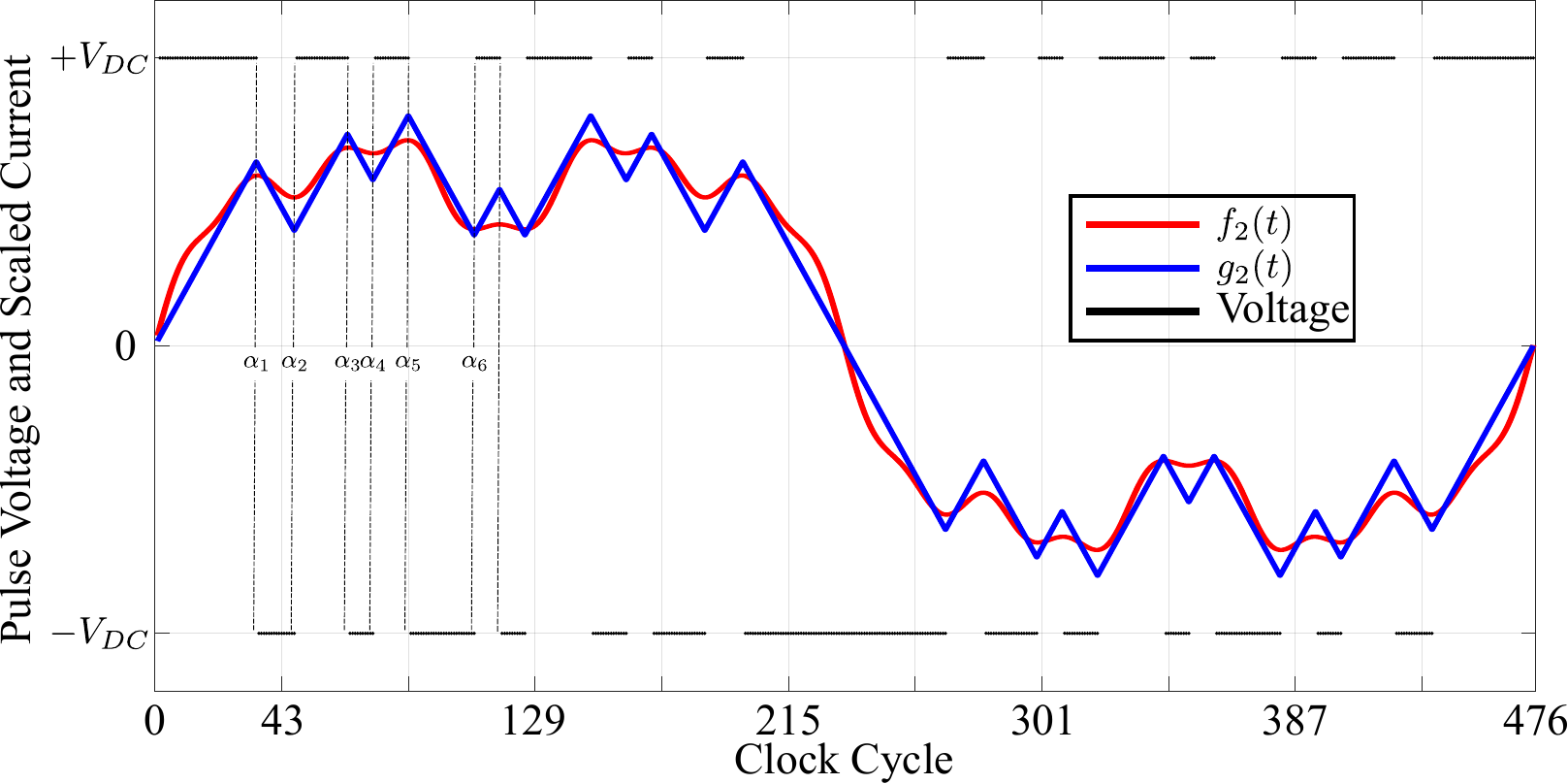}
\par\end{centering}
\caption{$f_{2}(t)$, $g_{2}(t)$ and the pulse voltage that generates $g_{2}(t)$\label{fig:Plots-of-f2}}
\end{figure}

Determining the number of output voltage levels of the amplifier is
another trade-off. As shown in \Figref{Plots-of-f2}, the resemblance
function $g_{2}(t)$ generated by the bipolar voltage across the coil
is able to approximate the objective function $f_{2}(t)$ with a tolerable
error in practice. It will be justified through practical test that
the bipolar output voltage is capable of generating the coil currents
with low distortion. In addition, there are only six switching angles
$\alpha_{1}$, $\alpha_{2}$, $\alpha_{3}$, $\alpha_{4}$, $\alpha_{5}$
and $\alpha_{6}$ in a quarter period. The highest switching frequency
is close to the $17^{th}$ harmonic frequency implying that the magnitude
of the harmonic frequencies higher than the highest transmitting frequency
will be low. Therefore, a low-pass filter is no longer needed.

\paragraph{Setup of optimisation problem}

According to (\ref{eq:fft-multi-coefficient}) the Fourier coefficients
$b_{p}$ of $g_{2}(t)$ can be calculated as (\ref{eq:fft-c}). Cautions
should be made that there are two falling edges, i.e. $+V_{DC}$ to
$0$ and $0$ to $-V_{DC}$ at $\alpha_{1}$, $\alpha_{3}$, $\alpha_{5}$,
two raising edges, i.e. $-V_{DC}$ to $0$ and $0$ to $+V_{DC}$
at $\alpha_{2}$, $\alpha_{4}$, $\alpha_{6}$, and one falling edge,
i.e. $+V_{DC}$ to $0$ at $\frac{\pi}{2}$. Substituting (\ref{eq:fft-c})
in (\ref{eq:opt}) gives rise to the optimisation problem (\ref{eq:opt-1})
where the dc voltage $V_{DC}$ and coil inductance $L$ have been
cancelled. Additionally, because we would like $b_{p}$ to be inversely
proportional to $p$, (\ref{eq:opt-1}) is in fact normalised where
each residual in the objective function should have similar level
of magnitude.

\begin{multline}
b_{p}=-\frac{4V_{{\it DC}}}{p^{2}\pi\omega L}\left(2\sum_{q=1}^{6}(-1)^{q}\sin\left(p\alpha_{q}\right)-\sin\left(\frac{p\pi}{2}\right)\right),\\
p=1,3,5,\cdots\label{eq:fft-c}
\end{multline}

\begin{figure*}[tbh]
\begin{multline}
\min_{0\leq\alpha_{1}\leq\alpha_{2}\leq\cdots\leq\alpha_{6}\leq\frac{\pi}{2},m_{p}}\sum_{p=1,3,7,17}\lambda_{p}\left(\frac{8}{p\pi}\sum_{q=1}^{6}(-1)^{q}\sin\left(p\alpha_{q}\right)-\frac{4}{p\pi}\sin\left(\frac{p\pi}{2}\right)+m_{p}\right)^{2},\\
subject\,to\,\left|\frac{8}{p\pi}\sum_{q=1}^{6}(-1)^{q}\sin\left(p\alpha_{q}\right)-\frac{4}{p\pi}\sin\left(\frac{p\pi}{2}\right)\right|\leq\epsilon_{p},\,\forall\,p\in\left\{ 5,9,11,13,15\right\} \label{eq:opt-1}
\end{multline}
\end{figure*}

The initial values of the six switching angles are the points where
the gradient of $f_{2}(t)$ is zero as shown in \Figref{Plots-of-f2}.
The angles can only have discrete values and the resolution is determined
by the product of the order of the highest harmonic and the ratio
of the clock frequency over the highest transmitting frequency. As
a result, the length of the lookup table which stores the control
sequence is $476$ as a product of $17$ and $28$ and hence the resolution
of angle is $\frac{2\pi}{476}\,rads$. Since every switching angle
is an integer multiplied by the resolution, it is convenient to use
the number of clock cycle to represent the angles. A set of initial
values is listed in the first row of \Tabref{Initial-values-and}.
These initial values are substituted into (\ref{eq:fft-c}) in order
to calculate the modulation index $m_{p}$. The objective $b_{p}$
of selected harmonics is inversely proportional to $p$, so the corresponding
modulation index $m_{p}$ should have the same value. As a result,
the average value of $m_{1}$, $m_{3}$, $m_{7}$ and $m_{17}$ when
substituting the initial angles is used as the initial value for $m_{1}$,
$m_{3}$, $m_{7}$ and $m_{17}$.

The constrain threshold $\epsilon_{p}$ is set up in a way that the
scaled modulation index $\frac{m_{p}}{p}$ for those undesirable harmonics
has the same bound. The modulation index $m_{p}$ is scaled by $p$
because we would like $b_{p}$ to have the same bound for undesirable
harmonics. All the weights $\lambda_{p}$ are configured as one for
the first setup and then the residual of the $17^{th}$ harmonic is
relaxed by setting $\lambda_{17}$ as zero which is the second setup.
The reason of such relaxation will be discussed in following paragraphs.
Details of the initial condition and two setups for the parameters
are listed in \Tabref{Initial-values-and}.

\paragraph{Results}

\begin{figure}[tbh]
\begin{centering}
\includegraphics[width=1\columnwidth]{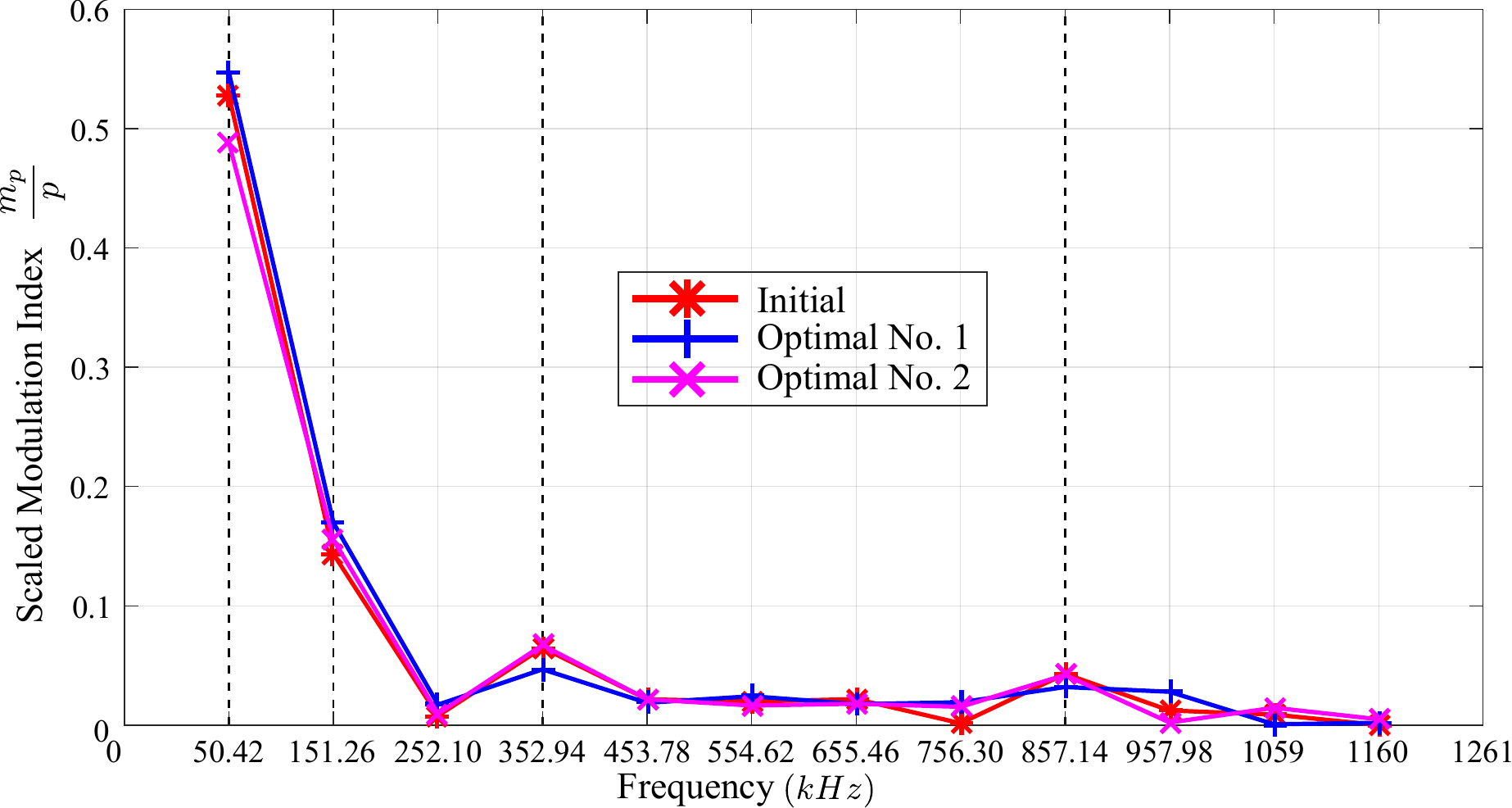}
\par\end{centering}
\caption{Scaled modulation index $\frac{m_{p}}{p}$ with respect to frequency
for the initial switching angles, optimal angles of the first setup
and the second setup of experiment No.1\label{fig:Plot-of-m}}
\end{figure}

The optimisation problem (\ref{eq:opt-1}) given the initial condition
under two setups of parameters is solved using the interior-point
algorithm. The resultant switching angles, total harmonic distortion
(THD) and scaled modulation index $\frac{m_{p}}{p}$ are listed in
\Tabref{Initial-values-and} where the THD is defined as (\ref{eq:THD}).
Here we use $\frac{m_{p}}{p}$ to represent $b_{p}$ because the former
is independent of the coil inductance $L$ while the latter is not.
An Fast Fourier Transform (FFT) is conducted to the resemblance functions
when the initial switching angles, the optimal angles of the first
setup and of the second setup are applied. The results are plotted
in \Figref{Plot-of-m}.

\begin{equation}
THD=\frac{\sqrt{\sum_{p\in\Psi}b_{p}^{2}}}{\sqrt{\sum_{p\in\Phi}b_{p}^{2}}}\times100\,\%\label{eq:THD}
\end{equation}

\begin{table*}[tbh]
\caption{Residual weight $\lambda_{p}$, constrain threshold $\epsilon_{p}$,
switching angle $\alpha_{q}$, scaled modulation index $\frac{m_{p}}{p}$
and the total harmonic distortion for the initial and two optimisation
setups of experiment No.1\label{tab:Initial-values-and}}

\resizebox{2.05\columnwidth}{!}{
\begin{centering}
\begin{tabular}{ccccccccccccccccccccc}
\toprule 
\multirow{2}{*}{} & \multirow{2}{*}{$\lambda_{1}$} & \multirow{2}{*}{$\lambda_{3}$} & \multirow{2}{*}{$\lambda_{7}$} & \multirow{2}{*}{$\lambda_{17}$} & \multirow{2}{*}{$\epsilon_{5}$} & \multirow{2}{*}{$\epsilon_{9}$} & \multirow{2}{*}{$\epsilon_{11}$} & \multirow{2}{*}{$\epsilon_{13}$} & \multirow{2}{*}{$\epsilon_{15}$} & \multicolumn{6}{c}{Switching angle (clock cycle)} & \multicolumn{4}{c}{$\frac{m_{p}}{p}$} & \multirow{2}{*}{$THD\,(\%)$}\tabularnewline
\cmidrule{11-20} \cmidrule{12-20} \cmidrule{13-20} \cmidrule{14-20} \cmidrule{15-20} \cmidrule{16-20} \cmidrule{17-20} \cmidrule{18-20} \cmidrule{19-20} \cmidrule{20-20} 
 &  &  &  &  &  &  &  &  &  & $\alpha_{1}$ & $\alpha_{2}$ & $\alpha_{3}$ & $\alpha_{4}$ & $\alpha_{5}$ & $\alpha_{6}$ & $50.42\,kHz$ & $151.26\,kHz$ & $352.94\,kHz$ & $857.14\,kHz$ & \tabularnewline
\midrule 
Initial & $-$ & $-$ & $-$ & $-$ & $-$ & $-$ & $-$ & $-$ & $-$ & $36$ & $49$ & $68$ & $77$ & $88$ & $111$ & $0.534$ & $0.145$ & $0.066$ & $0.043$ & $7.33$\tabularnewline
No. 1 & $1$ & $1$ & $1$ & $1$ & $0.10$ & $0.18$ & $0.22$ & $0.26$ & $0.30$ & $39$ & $51$ & $70$ & $82$ & $90$ & $111$ & $0.553$ & $0.171$ & $0.047$ & $0.032$ & $9.04$\tabularnewline
No. 2 & $1$ & $1$ & $1$ & $0$ & $0.10$ & $0.18$ & $0.22$ & $0.26$ & $0.30$ & $35$ & $48$ & $65$ & $74$ & $86$ & $111$ & $0.493$ & $0.157$ & $0.068$ & $0.043$ & $7.80$\tabularnewline
\bottomrule
\end{tabular}
\par\end{centering}
}
\end{table*}

\begin{figure}[tbh]
\begin{centering}
\includegraphics[width=1\columnwidth]{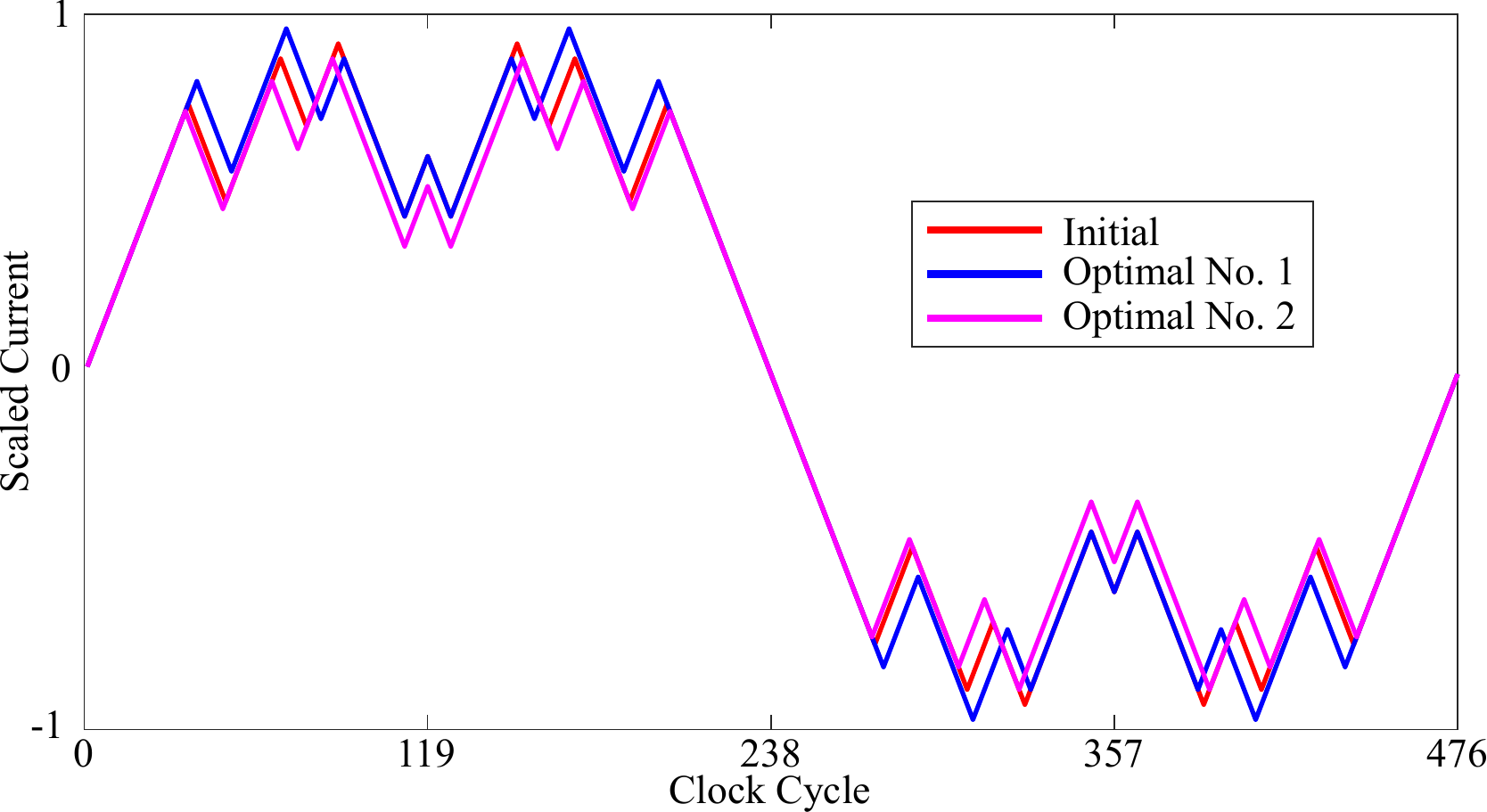}
\par\end{centering}
\caption{Waveform of the resemblance function for the initial switching angles,
optimal angles of the first setup and the second setup of experiment
No.1\label{fig:Plots-of-the}}
\end{figure}

It can been seen from \Tabref{Initial-values-and} that the initial
switching angles already lead to a low THD of $7.33\,\%$. However,
the power distribution among the four selected frequencies is not
the same as desired. The fourth component is higher while the second
and third component are lower than they should be. This situation
is improved by the optimisation under the first setup. Equal weights
are assigned to the four residuals resulting in a more desirable power
distribution. However, the THD has been increased to $9.04\,\%$ which
is mainly caused by the increase of the adjacent harmonic components
of $857.14\,kHz$ that can be seen in \Figref{Plot-of-m}. The reason
behind this is that initially the frequency of the highest-frequency
ripple is close to $857.14\,kHz$ which in turn gives rise to the
large component of $857.14\,kHz$ and a clean spectrum higher than
$857.14\,kHz$. The optimisation process tends to penalise the residual
of $857.14\,kHz$ by, unfortunately, increasing its adjacent harmonic
components rather than the lower desirable harmonics. Consequently,
it is reasonable to always relax the penalty of the '$benign\,ripple$'
as has been done in the second setup. As shown in \Tabref{Initial-values-and},
the scaled modulation index of the first three frequencies have a
better ratio comparing to those of the first optimisation while the
highest-frequency $\frac{m_{p}}{p}$ and THD are similar. The waveforms
of the associated resemblance functions for the three sets of switching
angles are shown in \Figref{Plots-of-the}.

\begin{figure}[tbh]
\begin{centering}
\includegraphics[width=1\columnwidth]{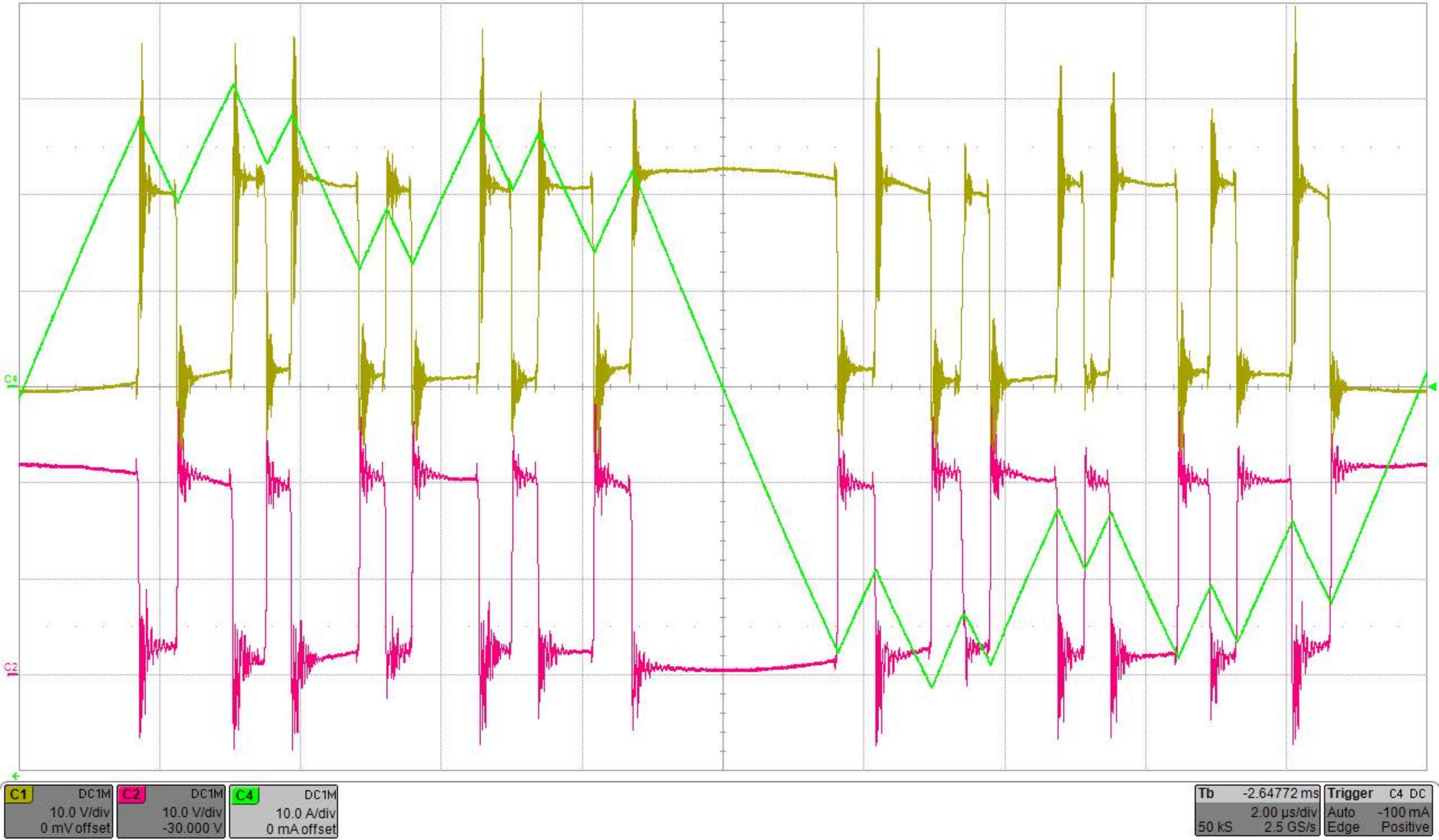}
\par\end{centering}
\caption{Coil current (green trace), output voltage of one half bridge (yellow
trace) and of the other half bridge (red trace) for optimisation No.
2 of experiment No. 1\label{fig:Coil-current-(green}}
\end{figure}

The optimal switching angles of the second optimisation setup are
implemented in the experimental system to drive a coil with an inductance
of about $1.4\,\mu H$. The screenshot of an oscilloscope displaying
the coil current and bipolar output voltage is shown in \Figref{Coil-current-(green}.
The peak-to-peak value of the current through the coil is higher than
$60\,A$. An FFT is done to the current and the largest ten root-mean-square
(RMS) values and the corresponding frequencies are listed in \Tabref{FFT-results-of}.

\begin{table}[tbh]
\caption{FFT results of the coil current for optimisation No. 2 of experiment
No. 1\label{tab:FFT-results-of}}

\resizebox{\columnwidth}{!}{
\begin{centering}
\begin{tabular}{cccccc}
\toprule 
No. & Frequency & Current (RMS) & No. & Frequency & Current (RMS)\tabularnewline
\midrule 
$1$ & $50.42\,kHz$ & $16.97\,A$ & $6$ & $554.62\,kHz$ & $0.76\,A$\tabularnewline
$2$ & $151.26\,kHz$ & $6.15\,A$ & $7$ & $252.10\,kHz$ & $0.49\,A$\tabularnewline
$3$ & $352.94\,kHz$ & $2.08\,A$ & $8$ & $1462.18\,kHz$ & $0.49\,A$\tabularnewline
$4$ & $857.14\,kHz$ & $1.49\,A$ & $9$ & $756.30\,kHz$ & $0.32\,A$\tabularnewline
$5$ & $655.46\,kHz$ & $0.81\,A$ & $10$ & $453.78\,kHz$ & $0.31\,A$\tabularnewline
\bottomrule
\end{tabular}
\par\end{centering}
}
\end{table}

\subsubsection{Experiment No. 2}

Compared to the first experiment, the objective of the second experiment
is to drive a higher-impedance coil in a lower spectrum from about
$1\,kHz$ to $200\,kHz$. The axis of the coil aperture would point
to the test samples. Application of this spectrum can be found in
the inspection of welding process, crack detection of metal samples
and thickness measurement of metal film, etc.

\begin{figure}[tbh]
\begin{centering}
\includegraphics[width=1\columnwidth]{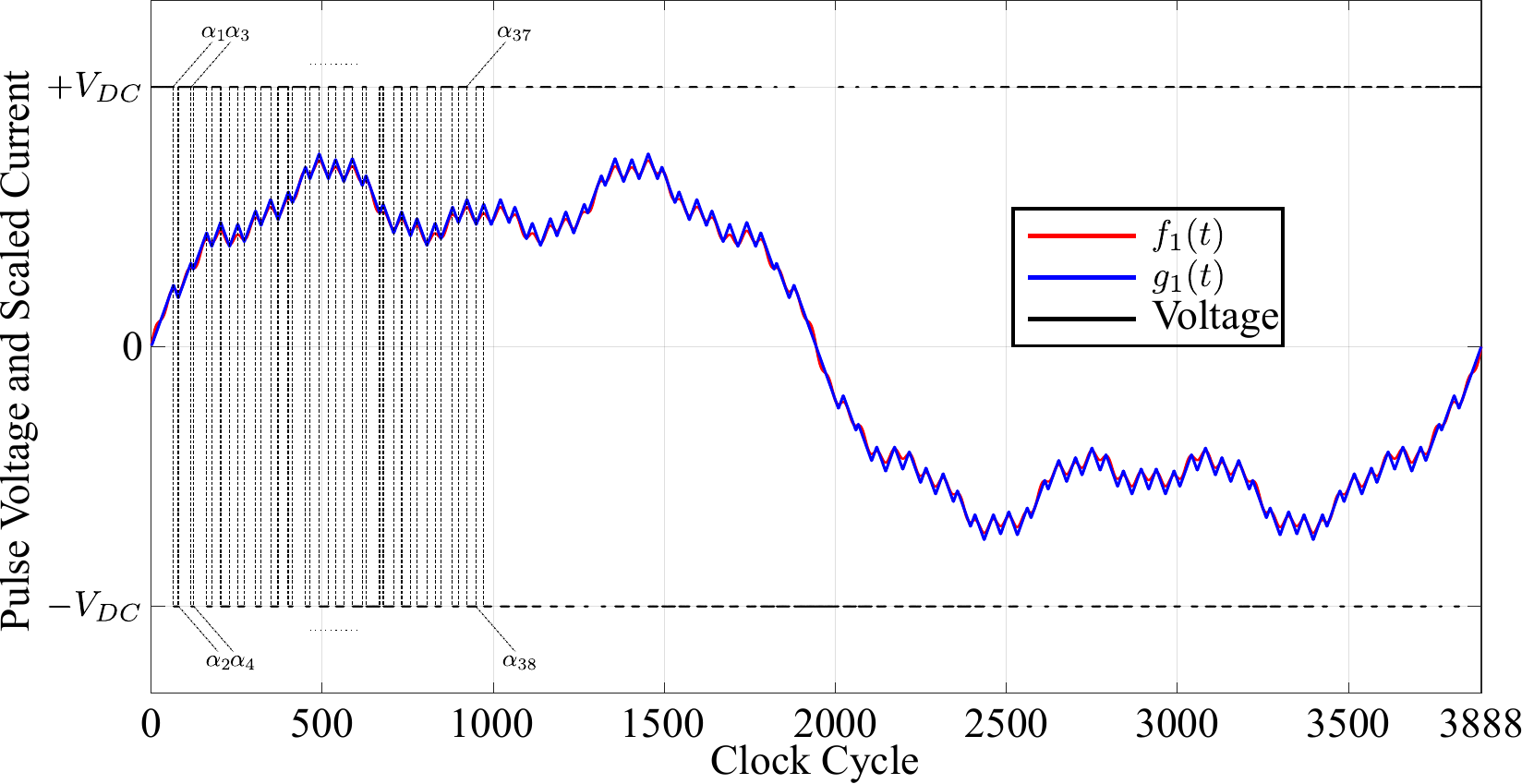}
\par\end{centering}
\caption{$f_{1}(t)$, $g_{1}(t)$ and the pulse voltage that generates $g_{1}(t)$\label{fig:Plots-of-f1}}
\end{figure}

The objective function $f_{1}(t)$, as shown in \Figref{Plots-of-f1},
is implemented in this experiment where the fundamental, $3^{rd}$,
$9^{th}$, $27^{th}$ and $81^{st}$ harmonics are involved. The resemblance
function $g_{1}(t)$ is also plotted in \Figref{Plots-of-f1} and
there are $38$ switching angles in a quarter period of $g_{1}(t)$.
Because the highest order harmonic is the $81^{st}$ harmonic, $g_{1}(t)$
requires more switching angles than $g_{2}(t)$. The initial values
of the switching angles are determined in the same way as it is in
the first experiment, i.e. examining the gradient of objective function.
The clock frequency is configured as $6.25\,MHz$ and the five transmitting
frequencies are chosen as about $1.61\,kHz$, $4.82\,kHz$, $14.47\,kHz$,
$43.40\,kHz$ and $130.21\,kHz$. As a result, the memory needs $3888\,bits$
to store the control sequence.

\begin{figure}[tbh]
\begin{centering}
\includegraphics[width=1\columnwidth]{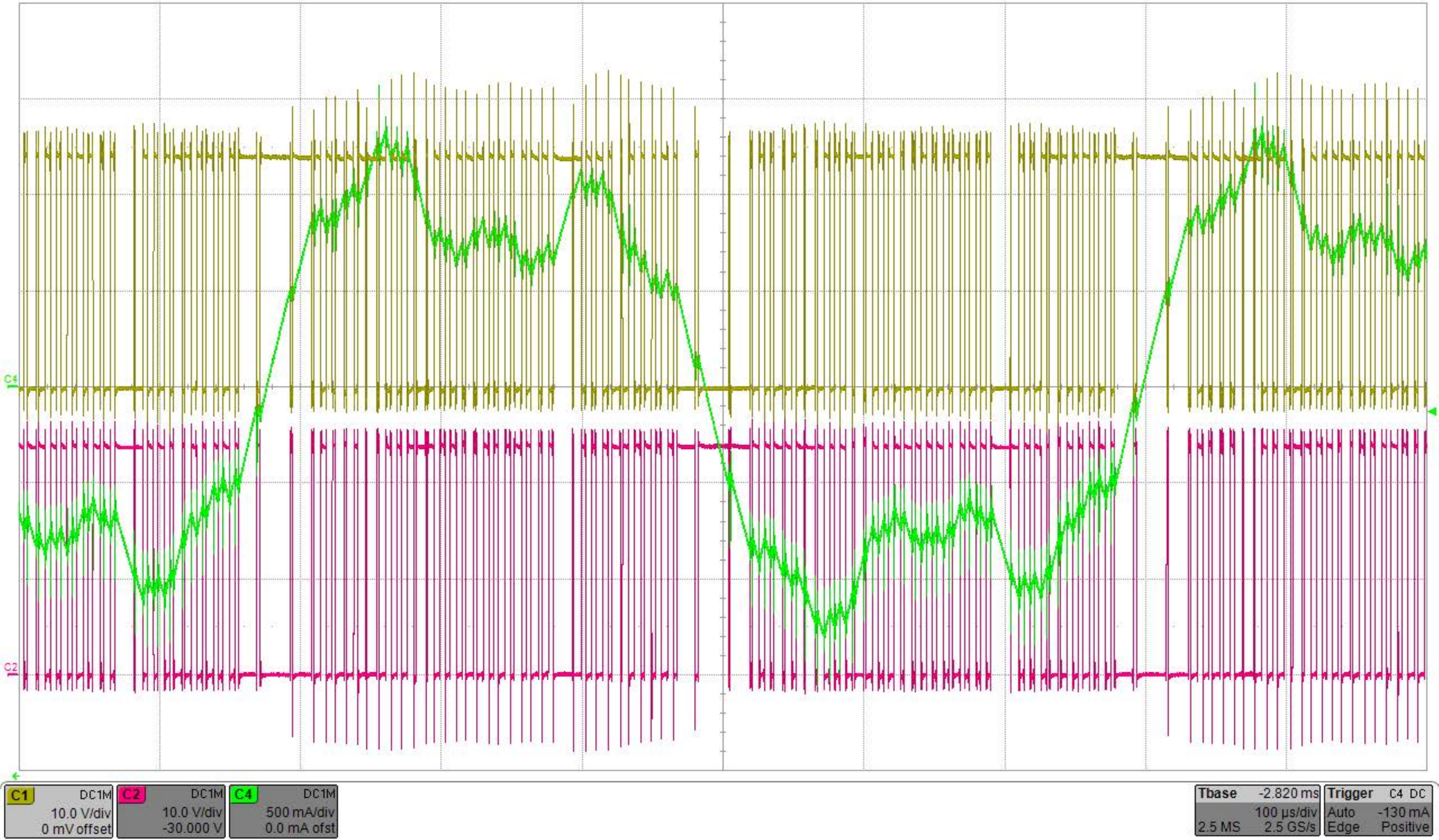}
\par\end{centering}
\caption{Coil current (green trace), output voltage of one half bridge (yellow
trace) and of the other half bridge (red trace) for experiment No.
2\label{fig:Coil-current-(green-1}}
\end{figure}

In order to show that a fast design could be done without the optimisation
process discussed in the first experiment, the initial values of the
$38$ switching angles are directly implemented in the experimental
system. The screenshot of an oscilloscope displaying the coil current
and bipolar output voltage is shown in \Figref{Coil-current-(green-1}.
The peak-to-peak value of the current through the coil is about $2.5\,A$.
An FFT is done to the current and the largest ten root-mean-square
(RMS) values and the corresponding frequencies are listed in \Tabref{FFT-results-of-1}.

\begin{table}[tbh]
\caption{FFT results of the coil current for experiment No. 2\label{tab:FFT-results-of-1}}

\resizebox{\columnwidth}{!}{
\begin{centering}
\begin{tabular}{cccccc}
\toprule 
No. & Frequency & Current (RMS) & No. & Frequency & Current (RMS)\tabularnewline
\midrule 
$1$ & $1.61\,kHz$ & $751.46\,mA$ & $6$ & $8.04\,kHz$ & $10.53\,mA$\tabularnewline
$2$ & $4.82\,kHz$ & $284.49\,mA$ & $7$ & $20.90\,kHz$ & $10.30\,mA$\tabularnewline
$3$ & $14.47\,kHz$ & $89.05\,mA$ & $8$ & $11.25\,kHz$ & $7.75\,mA$\tabularnewline
$4$ & $43.40\,kHz$ & $24.60\,mA$ & $9$ & $17.68\,kHz$ & $7.52\,mA$\tabularnewline
$5$ & $130.21\,kHz$ & $22.80\,mA$ & $10$ & $27.33\,kHz$ & $5.49\,mA$\tabularnewline
\bottomrule
\end{tabular}
\par\end{centering}
}
\end{table}

\subsubsection{Power and efficiency comparison between two experiments}

According to (\ref{eq:drainefficiency}), (\ref{eq:pout}), (\ref{eq:engcvfctr})
and (\ref{eq:qout}), the power and efficiency metrics of the two
experiments are calculated and listed in \Tabref{Efficiency}. In
terms of the efficiency $\eta$, the first experiment has a lower
value of 20.02\%, which is due to the fact that the coil has a high
quality factor and much of the power is dissipated by the MOSFETs.
In contrast, the efficiency $\eta$ is 76.92\% in the second experiment,
because the coil is lossy. The energy conversion factor $\zeta$ is
larger than unity for both experiments. The reactive power $Q_{out}$
for the first experiment is as large as 224.81 var with a total dissipated
power of 34.32 W. The second experiment has a lower $\zeta$ mainly
because of its higher impedance of the coil. If possible, we would
pursue a low efficiency $\eta$, because power dissipation in the
load would be a waste of energy. On the other hand, a large energy
conversion factor $\zeta$ is preferred, which implies that an ac
magnetic field would be generated while consuming low dc power.

\begin{table}[tbh]
\caption{Power and efficiency metrics of two experiments.\label{tab:Efficiency}}

\resizebox{\columnwidth}{!}{
\begin{centering}
\begin{tabular}{cccccccc}
\toprule 
Experiment & $V_{DC}$ & $I_{DC}$ & $P_{DC}$ & $P_{out}$ & $Q_{out}$ & $\eta$ & $\zeta$\tabularnewline
\midrule 
No. 1 & 24.00 V & 1.43 A & 34.32 W & 6.87 W & 224.81 var & 20.02\% & 6.55\tabularnewline
No. 2 & 24.00 V & 0.13 A & 3.12 W & 2.40 W & 8.62 var & 76.92\% & 2.76\tabularnewline
\bottomrule
\end{tabular}
\par\end{centering}
}
\end{table}

\section{Conclusions}

The instrumental development of MECT entailed a highly efficient amplifier
in the transmitting power stage. Switch-mode power amplifiers exhibited
compelling advantages in terms of efficiency. This article attempted
to serve as a bridge that introduced a switch-mode power amplifier
into the design of MECT instrumentation. Being able to transmit MSF
magnetic fields was critical in MECT but was not a primary goal in
the research of modulation strategy of switch-mode amplifiers or converters.
Hence, this article also examined the MSF capability of the SHEPWM
technique. A Class D power amplifier was developed in this article
aiming at generating high MSF currents in the transmitting coil. An
MSF-SHEPWM method was proposed to modulate the pulse voltage across
the coil. Three concise steps could summarize the new modulation strategy,
i.e., time-domain synthesis of harmonic signals, frequency-domain
analysis of the resemblance signal, and optimisation of the resultant
Fourier coefficients. An FPGA-based experimental system was developed,
and the proposed methodology was implemented in two experiments. The
first experiment transmitted magnetic fields exciting at four simultaneous
frequencies with a THD as low as 8\%. The peak-to-peak value of the
coil current was as high as 60A while the total power dissipation
was low.The size of the circuit board was small and no large heat
dissipating facilities were needed. The second experiment tries to
illustrate a fast design process showing that the proposed method
was still effective even with the first step only. Five frequencies
were transmitted in a smaller span of spectrum comparing to the first
experiment. The two experiments have shown that the design is versatile
and could be applied to different electromagnetic inductive sensing
scenarios.

\section*{Acknowledgement}

All research data supporting this publication are directly available
within this publication.

\bibliographystyle{IEEEtranTIE}
\bibliography{REF_19-TIE-1030}

\begin{IEEEbiography}[{\includegraphics[width=1in,height=1.25in]{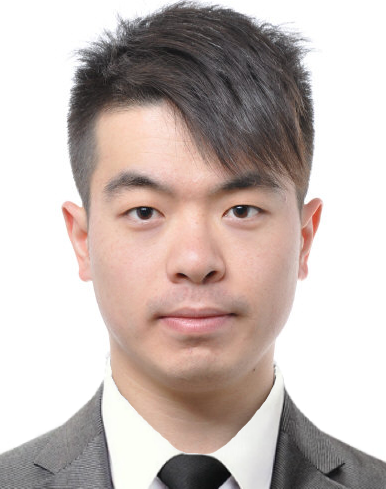}}]{Yang Tao}
(M\textquoteright 19) received the B.Sc. degree in automation and
the M.Sc. degree in control science and engineering from Tianjin University,
Tianjin, China, in 2010 and 2013, respectively, and the Ph.D. degree
in electrical and electronic engineering from the University of Manchester,
Manchester, U.K., in 2018.

He is currently an Research Associate with the Department of Electrical
and Electronic Engineering, School of Engineering, University of Manchester.
His research interests include electromagnetic instrumentation, metal
detection, multifrequency power inverter, electromagnetic tomography,
inverse problem, sparse representation, and deep learning. 
\end{IEEEbiography}

\begin{IEEEbiography}[{\includegraphics[width=1in,height=1.25in]{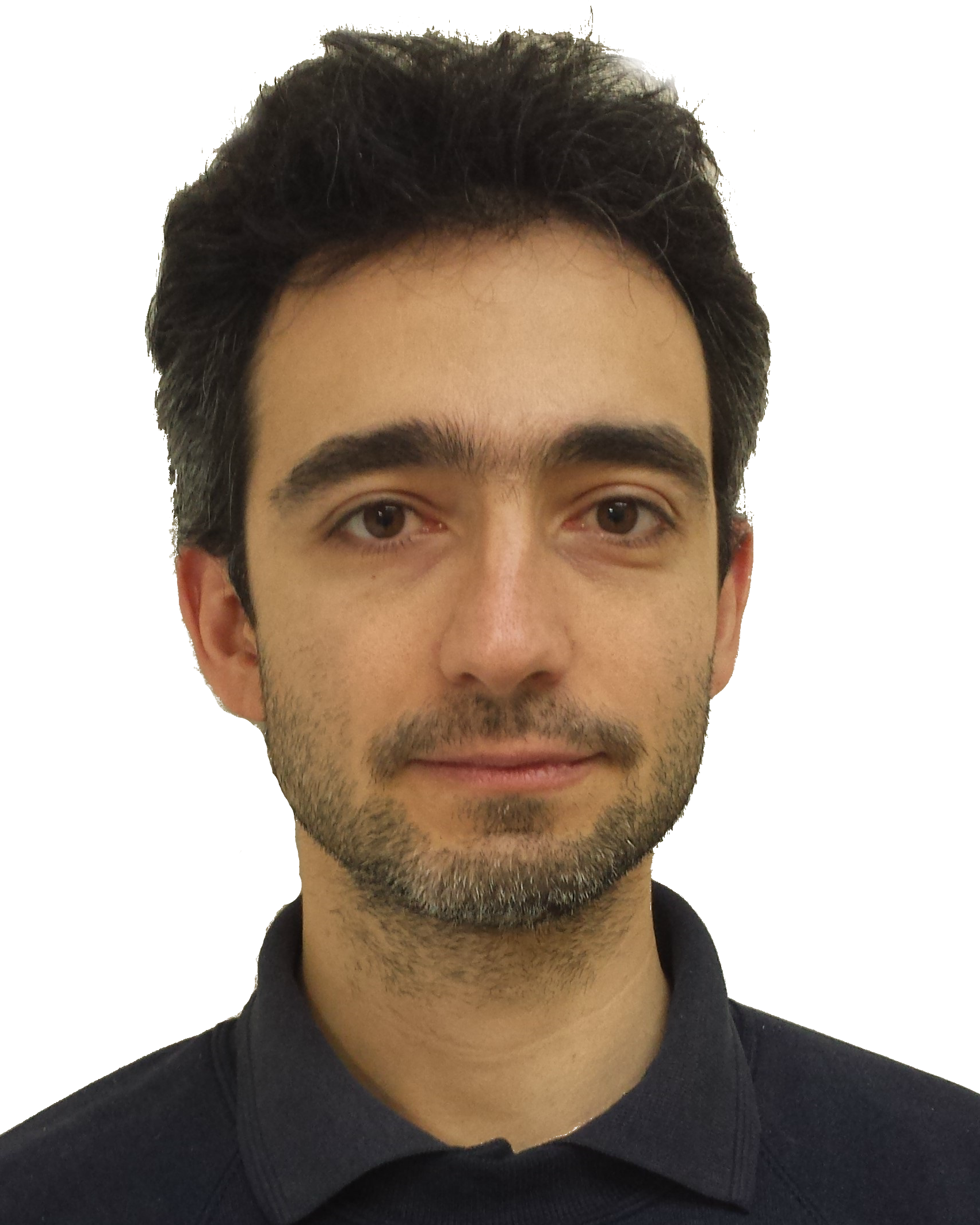}}]{Christos Ktistis}
received the B.Sc. degree in mechanical engineering from the Technological
Educational Institution, Serres, Greece, in 2000, the M.Sc. degree
in mechatronics from Lancaster University, Lancaster, U.K., in 2002,
and the Ph.D. degree in electrical and electronic engineering from
The University of Manchester, Manchester, U.K., in 2007.

He is currently a Visitor with the Department of Electrical and Electronic
Engineering, School of Engineering, The University of Manchester.
\end{IEEEbiography}

\begin{IEEEbiography}[{\includegraphics[width=1in,height=1.25in]{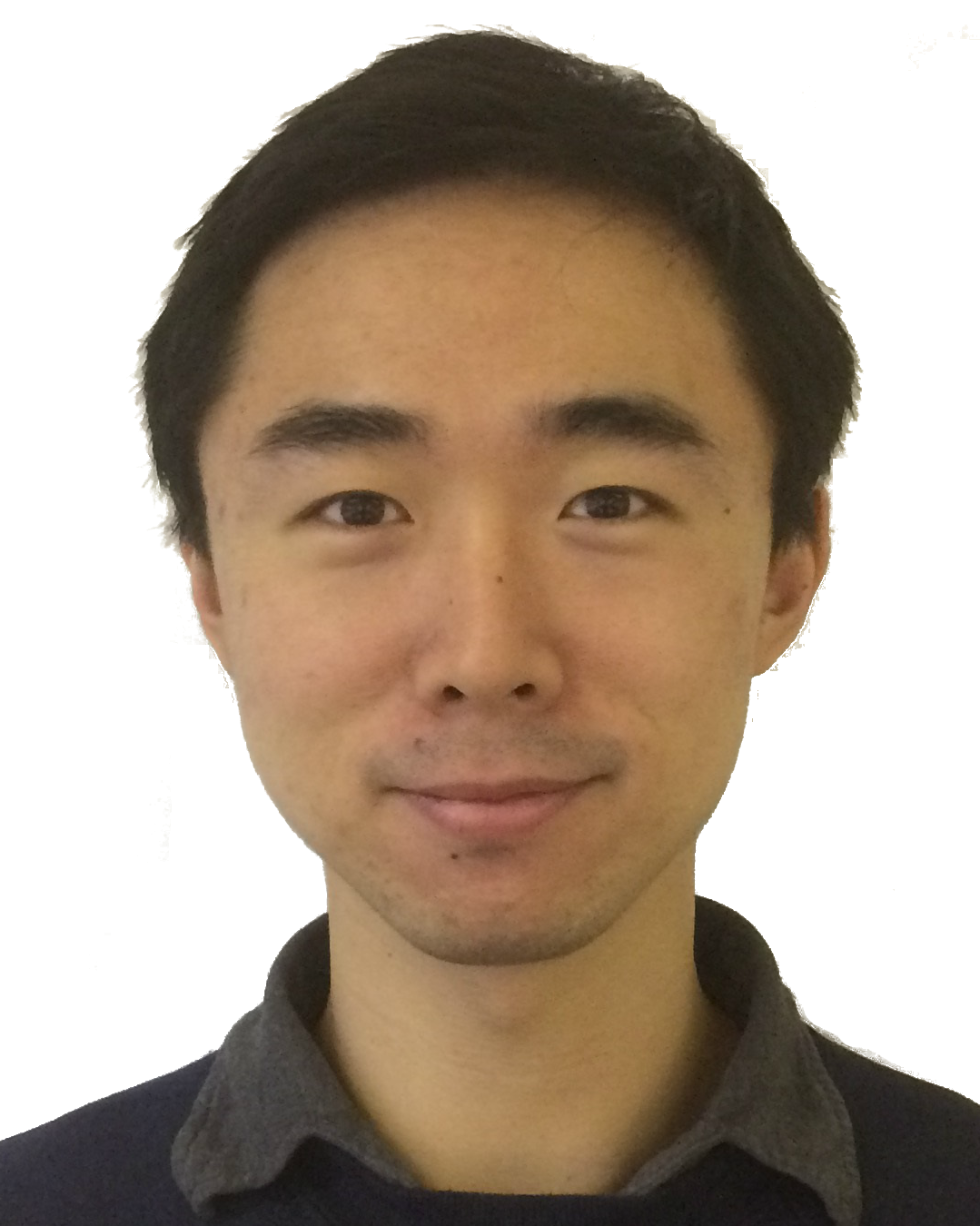}}]{Yifei Zhao}
received the B.Eng. (Hons.) degree and the Ph.D. degree in electrical
and electronic engineering from The University of Manchester, Manchester,
U.K., in 2010 and 2013, respectively.

He is currently a Visitor with the Department of Electrical and Electronic
Engineering, School of Engineering, The University of Manchester.
\end{IEEEbiography}

\begin{IEEEbiography}[{\includegraphics[width=1in,height=1.25in]{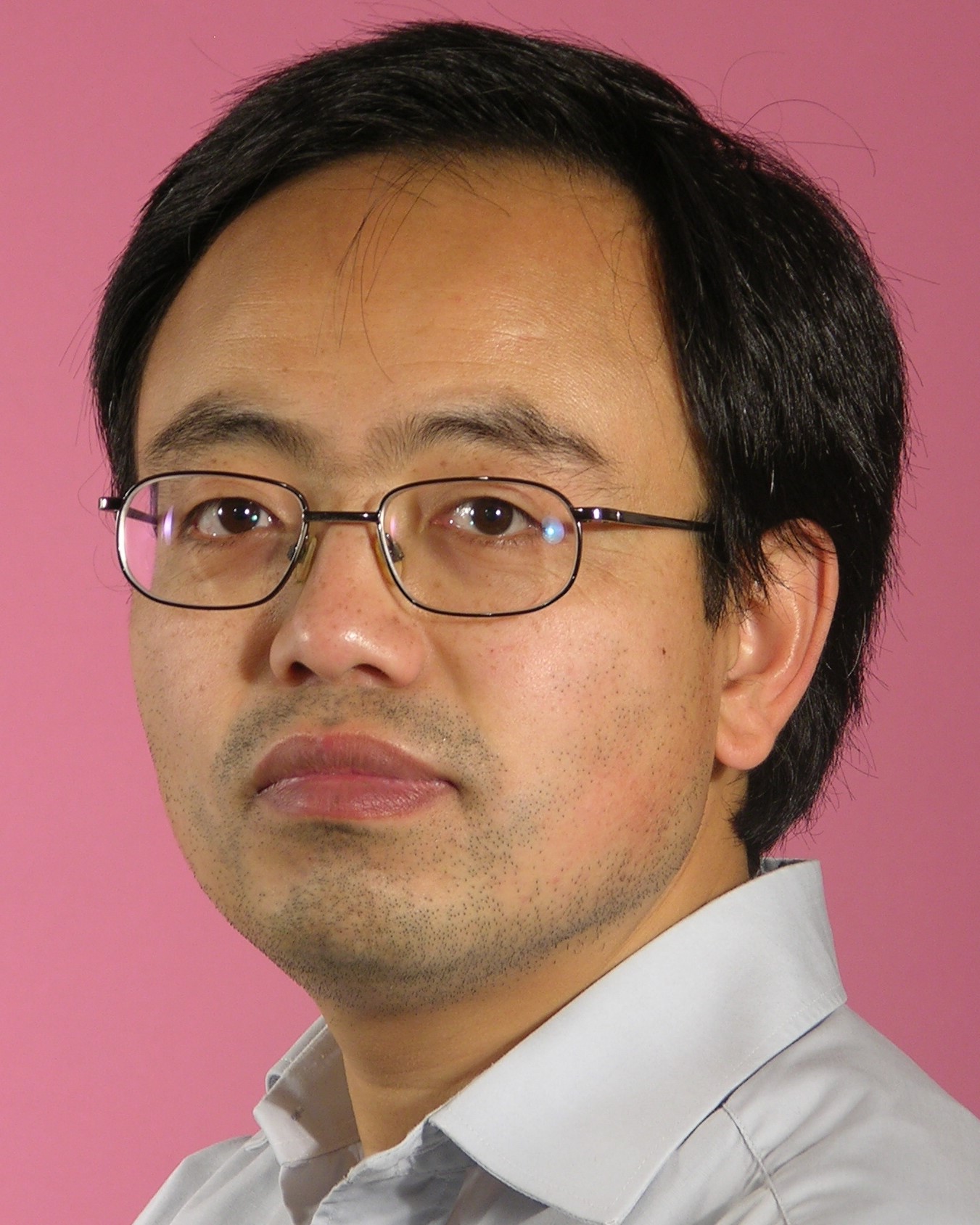}}]{Wuliang Yin}
 (M\textquoteright 05--SM\textquoteright 06) received the B.Sc.
and the M.Sc. degrees in electronic measurement and instrumentation
from Tianjin University, Tianjin, China, in 1992 and 1995, respectively,
and the Ph.D. degree in automotive electronics from Tsinghua University,
Beijing, China, in 1999.

He was appointed as a Mettler Toledo (MT) Sponsored Lecturer with
the Department of Electrical and Electronic Engineering, School of
Engineering, The University of Manchester, Manchester, U.K., in 2012,
and was promoted to a Senior Lecturer in 2016. He has authored 1 book,
more than 230 papers, and was granted more than 10 patents in the
area of electromagnetic sensing and imaging.

Dr. Yin was a recipient of the 2014 and 2015 Williams Award from the
Institute of Materials, Minerals and Mining and the Science and Technology
Award from the Chinese Ministry of Education in 2000. 
\end{IEEEbiography}

\begin{IEEEbiography}[{\includegraphics[width=1in,height=1.25in]{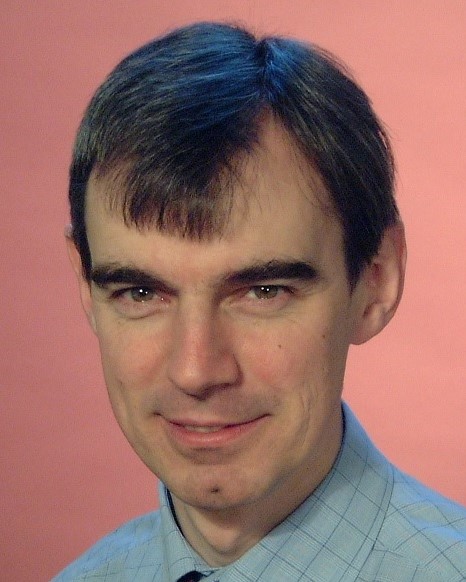}}]{Anthony J. Peyton}
 received the B.Sc. degree in electrical engineering and electronics
and the Ph.D. degree in medical instrumentation from The University
of Manchester Institute of Science and Technology (UMIST), Manchester,
U.K., in 1983 and 1986, respectively.

He was appointed as a Principal Engineer with Kratos Analytical Ltd.
in 1989, developing precision electronic instrumentation systems for
magnetic sector and quadrupole mass spectrometers, from which an interest
in electromagnetic instrumentation was developed. He returned to UMIST
as a Lecturer and worked with the Process Tomography Group. He moved
to Lancaster University in 1996 taking up the post of Senior Lecturer
and promoted to a Reader in Electronic Instrumentation in July 2001
and a Professor in May 2004. Since December 2004, he has been a Professor
of Electromagnetic Tomography Engineering with the University of Manchester,
Manchester. He has been a Principal Investigator of numerous national-
and industry-funded projects and a partner of ten previous EU projects,
one as a coordinator. He has been a coauthor on almost 140 international
journal papers, two books, several hundred conference papers, and
12 patents in areas related to electromagnetics and tomography. His
main research interests include instrumentation, applied sensor systems,
and electromagnetics. 
\end{IEEEbiography}

\end{document}